\documentclass[review]{elsarticle}

\usepackage{lineno}
\modulolinenumbers[5]
\usepackage{commath}
\usepackage{soul}
\usepackage{amsmath}
\usepackage[RPvoltages,europeanresistors,americaninductors]{circuitikz}
\usepackage{amssymb}
\usepackage{multicol,lipsum}
\usepackage{stfloats}

\usepackage{booktabs}
\usepackage{flushend}
\usepackage{setspace}
\usepackage{lscape}

\newtheorem{thm}{Theorem}

\newdefinition{rmk}{Remark}
\newproof{pf}{Proof}
\newdefinition{definition}{Definition}

\begin{document}

\begin{frontmatter}

\title{Cell-Level State of Charge Estimation for Battery Packs Under Minimal Sensing}


\author[OU_add]{Dong Zhang}
\author[ULB_add]{Luis D. Couto}
\author[Ox_add]{Ross Drummond}
\author[CMU_add]{Shashank Sripad}
\author[CMU_add]{Venkatasubramanian Viswanathan\corref{correspondingauthor}}
\cortext[correspondingauthor]{Corresponding author}
\ead{venkvis@cmu.edu}

\address[OU_add]{School of Aerospace and Mechanical Engineering, University of Oklahoma, 660 Parrington Oval, Norman, OK, USA}
\address[CMU_add]{Department of Mechanical Engineering, Carnegie Mellon University, 5000 Forbes Avenue, Pittsburgh, PA, USA}
\address[ULB_add]{Department of Control Engineering and System Analysis, Universit\'e Libre de Bruxelles, B-1050 Brussels, Belgium}
\address[Ox_add]{Department of Engineering Science, University of Oxford, 17 Parks Road, OX1 3PJ, Oxford, United Kingdom}

\begin{abstract}
This manuscript presents an algorithm for individual Lithium-ion (Li-ion) battery cell state of charge (SOC) estimation in a large-scale battery pack under minimal sensing, where only pack-level voltage and current are measured. For battery packs consisting of up to thousands of cells in electric vehicle or stationary energy storage applications, it is desirable to estimate individual cell SOCs without cell local measurements in order to reduce sensing costs. Mathematically, pure series connected cells yield dynamics given by ordinary differential equations under classical full voltage sensing. In contrast, parallel--series connected battery packs are evidently more challenging because the dynamics are governed by a nonlinear differential--algebraic equations (DAE) system. The majority of the conventional studies on SOC estimation for battery packs benefit from idealizing the pack as a lumped single cell which ultimately lose track of cell-level conditions and are blind to potential risks of cell-level over-charge and over-discharge. This work explicitly models a battery pack with high fidelity cell-by-cell resolution based on the interconnection of single cell models, and examines the observability of cell-level state with only pack-level measurements. A DAE-based state observer with linear output error injection is formulated, where the individual cell SOC and current can be reconstructed from minimal number of pack sensing. The mathematically guaranteed asymptotic convergence of differential and algebraic state estimates is established by considering local Lipschitz continuity property of system nonlinearities. Simulation results for Graphite/NMC cells illustrate convergence for cell SOCs, currents, and voltages.
\end{abstract}


\end{frontmatter}


\section{Introduction}

\begin{figure}[t]
	\centering
	\includegraphics[trim = 0mm 0mm 0mm 0mm, clip, width=0.7\linewidth]{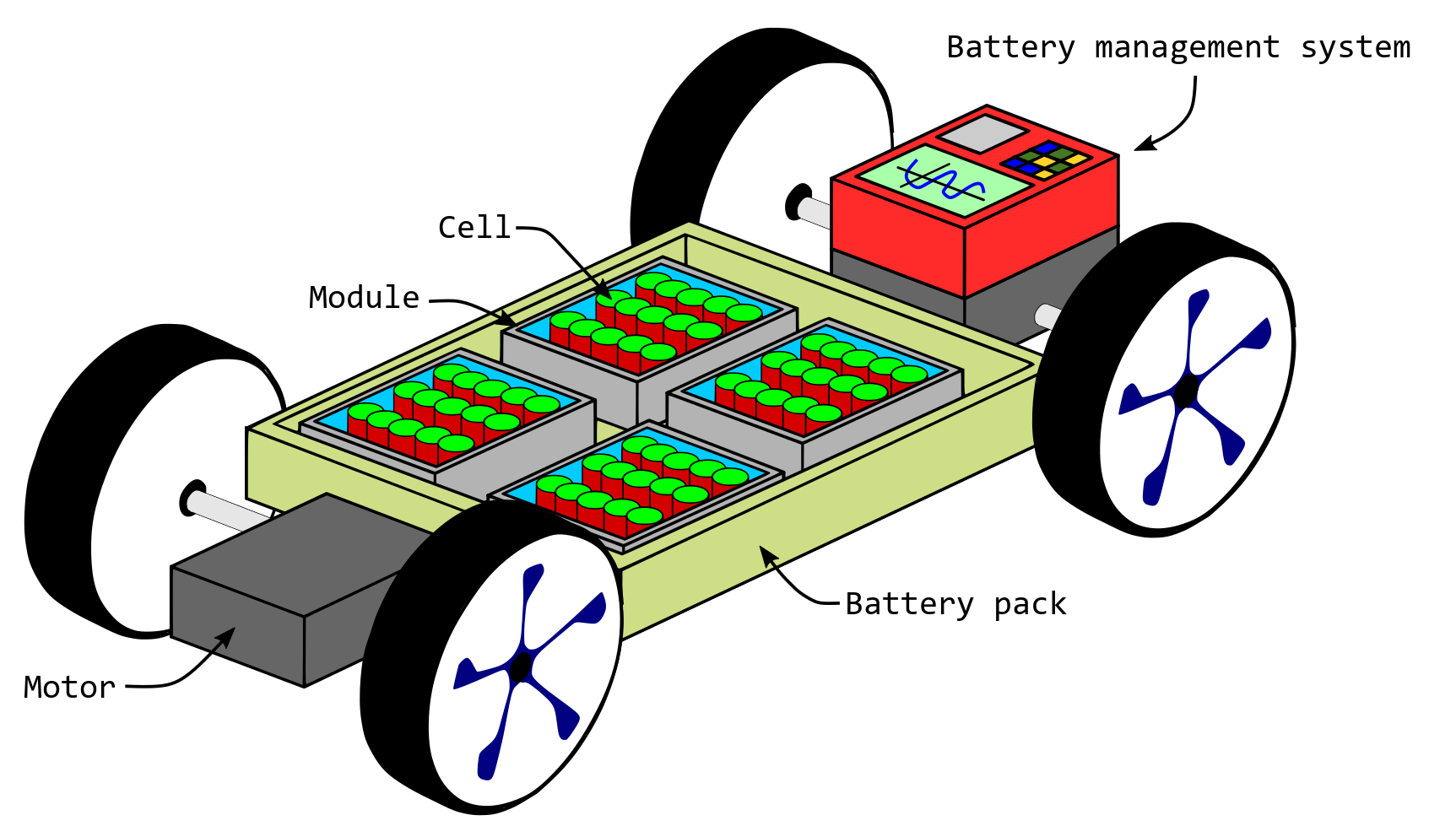}
	\caption{An illustrative schematic of a battery pack inside an electric vehicle with an on-board battery management system.}
	\label{fig:EV_Pack}
\end{figure}

Lithium-ion (Li-ion) batteries have emerged as one of the most prominent energy storage devices for large-scale energy applications, e.g., hybrid electric vehicles (HEV), pure electric vehicles (EV), and smart grids, due to their high energy and power density, low self-discharge, long lifetime, and rapidly falling prices \cite{chaturvedi2010algorithms,ziegler_re-examining_2021}. A battery pack system generally consists of hundreds or thousands of single cells connected via parallel and series connections in order to fulfill the requirements of high-energy and high-power applications \cite{Zhong-2014}. For example, the 75 kWh battery pack in Tesla Model 3 contains in total 4,416 Li-ion cylindrical 2170 cells organized in four modules (two 25 series and two 23 series) \cite{ulrich2020gm}. Each series component has 46 battery cells wired in parallel. See an illustrative example in Figure \ref{fig:EV_Pack} for an EV with a battery pack as well as an on-board battery management system (BMS). Meanwhile, it is also well-known that Li-ion cells in battery packs are sensitive to over-(dis)charge, high currents, and degrade over their lifetime \cite{Lin-2015}. Accurate estimation of the battery internal variables, such as state of charge (SOC) and local currents, enables a battery management system to ensure that individual in-pack cells do not violate safety constraints while prolonging battery service life via online control and cell-level health diagnosis.

Battery pack system modeling can be divided into three categories. The first approach treats the entire pack as one lumped single cell \cite{Castano-2015}. However, the internal states of individual cells within the pack are likely to be different. Results from \cite{baumhofer2014production,baumann2018parameter} show that there exists considerable cell-to-cell variability even for cells manufactured in the same batch.
Therefore, within a pack, some cells are more prone to violate safety-critical constraints than others, and such issues cannot be resolved from the lumped single cell approach. The second modeling approach also relies on a single cell model, but it focuses on specific in-pack cells -- the weakest and the strongest ones, as representatives of the pack dynamics \cite{Zhong-2014,Hua-2015}. Although convenient from the computational viewpoint, these extreme cells need to be identified and they may change with time, and this is generally not considered. The last approach is based on the interconnection of single cell models \cite{Zheng-2013,Zhang-2016,Zhao-2015}. This approach benefits from high fidelity cell-by-cell resolution, but it might suffer from high real-time computational burden. To counteract this computational challenge, most of these approaches resort to equivalent circuit models, which tend to have a low complexity when compared to more sophisticated electrochemical models.

Typical battery packs in EVs and storage applications, as mentioned previously, usually contain hundreds or thousands of cells. Conventionally, to carefully monitor every single cell with divergence of characteristics, voltage sensing is placed on each cell or every group of cells in parallel \cite{Lin-2015}. In practice, this task inevitably demands a significant amount of sensing hardware and labor, and is considered to be uneconomical. Moreover, as pointed out in \cite{kim2011efficient}, the failure rate of the battery system increases with more hardware components. Further, large-scale battery packs generate massive amount of data if a large number of voltage sensors are deployed, challenging the BMS's storage and computational capability limits \cite{zhou2020massive}. All these issues motivate a substantial reduction on number of sensing hardware.

In battery systems-and-control community, most of the existing studies on SOC estimation for battery packs resort to idealizing the entire battery pack as a lumped cell and define a pack-level SOC \cite{Zhong-2014,Z2016line,sepasi2014improved,wang2007combined,hu2010estimation}. However, such methods completely lose track of cell-level conditions and are blind to potential risks of cell-level over-charge and over-discharge. 
To date, the cell-level SOC estimation problem in a battery pack under reduced sensing has not been sufficiently explored. Series cell SOC estimation using only the total voltage measurement has been studied previously in \cite{Lin-2015,Couto_Series,couto2020estimation}, whereas the estimation for parallel (and parallel-series) configuration has been overlooked for multiple reasons. First, the cells in parallel are widely considered to behave as one single cell. However, an implicit assumption behind this reasoning is that the applied current is evenly split amongst the cells in parallel. This is rarely true in practice due to cell heterogeneity, such as non-uniform parameter values and temperatures \cite{Bruen-2016}. 
Secondly, the estimation problem for battery cells in series is arguably easier to solve than the parallel counterpart, because in the series case, the input current to each battery cell is the same and it can be practically measured. In the parallel and parallel-series cases, each cell's local current is unknown and determined by nonlinear algebraic constraints. Under reduced sensing scenario, the parallel configuration turns out to be a differential-algebraic equation (DAE) system that requires non-trivial state estimation theories \cite{ZhangDescriptor}.

A DAE system, \emph{a.k.a.} a descriptor or singular system, involving both differential and algebraic equations, is a powerful modeling framework that generalizes ordinary differential (normal) systems \cite{duan2010analysis}. 
The state observer design for linear descriptor systems is a rich research topic \cite{Niko-1992,Chisci-1992,Darouach-1995}. In contrast, state observers for nonlinear DAE systems is less prolific. Some relevant contributions encompass a local asymptotic state observer 
\cite{Boutayeb-1995}, looking at the system as set of differential equations on a restricted manifold 
\cite{Zimmer-1997}, and an index-1 DAE observer \cite{Aslund-2006}. 
Other works consider the case of Lipschitz nonlinearities \cite{rajamani1998observers}, which have served as a basis for Lyapunov-based observer design using the linearized system \cite{kaprielian1992observer}, and linear matrix inequality (LMI) approaches producing state observers in singular \cite{Lu-2006} and non-singular \cite{Darouach-2008} forms. Another Lipschitz system was considered in \cite{Shields-1997}, where the temporal separation between slow and fast dynamics was exploited to design a robust state observer. 
Nonlinear DAE systems have also been estimated through moving horizon approaches \cite{Albuquerque-1997} and Kalman filters \cite{Becerra-2001,Mandela-2010,Puranik-2012}.

In light of the aforementioned literature and research gaps, it is imperative to tackle the challenges in SOC estimation for individual cells in a battery pack under reduced sensing. In fact, we will explore the setting where only the overall pack voltage and current can be measured in real time denoted here as \textit{minimal sensing}, which will be showcased in Figure \ref{fig:Pack}. The proposed estimation approach contrasts with the majority of existing studies that lump the entire battery pack as one virtual cell, and others that propose the measurements of few specific cells inside a pack.
The proposed state estimator is realized by adopting an interconnection of single cell models with high fidelity cell-by-cell resolution. To the best of the authors’ knowledge, this is the first attempt in the literature to estimate the cell-level state in a battery pack with parallel-series configurations using only pack-level sensing, which is particularly challenging due to multiple technical reasons. First, the pack models with cell-level resolution are generally governed by complex differential-algebraic equations that require more sophisticated analysis and observer design theories than those for normal ordinary differential equation (ODE) systems. Second, under extremely limited measurable signals, namely only pack-level voltage and current, system observability is significantly deteriorated. This potentially renders locally unobservable conditions. Ultimately, this paper departs from previously existing works by
\begin{enumerate}[(1)]
    \item proposing a novel modeling framework for Li-ion battery packs with parallel-series connected heterogeneous cells as a nonlinear DAE (descriptor) system;
    \item rigorously analyzing the nonlinear local (smooth) observability conditions of such a nonlinear DAE system;
    \item designing a provably convergent Lyapunov-based state observer for the battery pack nonlinear DAE model to estimate cell-level SOCs and currents, utilizing only pack-level voltage and current measurements (minimal sensing).
\end{enumerate}
In particular, the system analysis and observer design proposed in this work is directly performed on the original nonlinear DAE model of the battery pack without any model reductions, thus retaining the physical significance of the equations and of the phenomena that they represent. This notably further separates our work from the majority of the existing efforts.

The reminder of this paper is organized as follows. Section \ref{s:mot} motivates the importance of monitoring cell-level SOC with the presence of heterogeneity among cells in a pack. Section \ref{s:mod} introduces the modeling framework for a battery pack with minimal sensing. Section \ref{s:obs_ana} provides the local observability analysis for the nonlinear battery pack DAE system. Section \ref{s:obs_design} proposes the state observer design and its asymptotic convergence analysis. Finally, the effectiveness of the proposed approach is illustrated in Section \ref{s:sim} via numerical simulations. Conclusions are drawn in Section \ref{s:conclusion}.

\textbf{Notation.} Throughout the work, the symbol $\mathbf{I}_{p \times q}$ denotes the identity matrix with dimension $p \times q$, and $\mathbf{0}_{p}$ indicates a zero column vector with dimension $p$. First-order time derivative of a variable $x$ is represented by $\dot{x}$, and high derivatives are denoted as $x^{(j)}$ where $j$ is the number of differentiation. $\mathbb{R}$ and $\mathbb{C}$ indicate the set of real numbers and complex numbers, respectively. Furthermore, a matrix $[A_1 \quad A_2]$ is 1-full if ${\rm rank}([A_1 \quad A_2]) = n + {\rm rank}(A_2)$,
where $A_1$ offers the first $n$ columns.

\section{Problem Formulation \& Motivation} \label{s:mot}

\begin{figure*}
\centering
\begin{circuitikz}[american voltages]

\draw (0,0) to [ammeter] (1.5,0);
\draw (1.5,0) to [short, *-] (1.5,1);
\draw (1.5,1) to [short, *-] (1.5,2);
\draw (1.5,0) to [short, *-] (1.5,-1);
\draw (1.5,-1) to [short, *-] (1.5,-2);
\draw (1.5,-1) to [R] (3.5,-1);
\draw (1.5,-2) to [R] (3.5,-2);
\draw (1.5,1) to [R] (3.5,1);
\draw (1.5,2) to [R] (3.5,2);
\draw (3.5,1) to [short, *-] (3.5,2);
\draw (3.5,0) to [short, *-] (3.5,1);
\draw (3.5,-1) to [short, *-] (3.5,0);
\draw (3.5,-1) to [short, *-] (3.5,-2);

\draw [dashed] (2.5,0.6) to (2.5,0) node[right]{$N_p$};
\draw [dashed] (2.5,0) to (2.5,-0.6);

\draw (3.5,0) to [short, *-] (4.5,0);
\draw (4.5,0) to [short, *-] (4.5,1);
\draw (4.5,1) to [short, *-] (4.5,2);
\draw (4.5,0) to [short, *-] (4.5,-1);
\draw (4.5,-1) to [short, *-] (4.5,-2);
\draw (4.5,-1) to [R] (6.5,-1);
\draw (4.5,-2) to [R] (6.5,-2);
\draw (4.5,1) to [R] (6.5,1);
\draw (4.5,2) to [R] (6.5,2);
\draw (6.5,1) to [short, *-] (6.5,2);
\draw (6.5,0) to [short, *-] (6.5,1);
\draw (6.5,-1) to [short, *-] (6.5,0)  node[left](p2){$ $}; 
\draw (6.5,-1) to [short, *-] (6.5,-2);

\draw [dashed] (5.5,0.6) to (5.5,0) node[right]{$N_p$};
\draw [dashed] (5.5,0) to (5.5,-0.6);

\draw [dashed] (p2) to (7.5,0) node[above]{$N_s$};
\draw [dashed] (7.5,0) to (8.5,0);

\draw (8.5,0) to [short, *-] (8.5,1);
\draw (8.5,1) to [short, *-] (8.5,2);
\draw (8.5,0) to [short, *-] (8.5,-1);
\draw (8.5,-1) to [short, *-] (8.5,-2);
\draw (8.5,-1) to [R] (10.5,-1);
\draw (8.5,-2) to [R] (10.5,-2);
\draw (8.5,1) to [R] (10.5,1);
\draw (8.5,2) to [R] (10.5,2);
\draw (10.5,1) to [short, *-] (10.5,2);
\draw (10.5,0) to [short, *-] (10.5,1);
\draw (10.5,-1) to [short, *-] (10.5,0)  node[left](p3){$ $}; 
\draw (10.5,-1) to [short, *-] (10.5,-2);

\draw [dashed] (9.5,0.6) to (9.5,0) node[right]{$N_p$};
\draw [dashed] (9.5,0) to (9.5,-0.6);

\draw (10.5,0) to [short, *-] (12,0);

\draw (0,0) -- (0,-3) to[voltmeter] (12,-3) -- (12,0);

\end{circuitikz}
\caption{Battery pack configuration under consideration. It consists of $N_s$ parallel modules in series, whereas each parallel module has $N_p$ cells connected in parallel. The $i$-th battery cell in the \mbox{$j$-th} parallel module is denoted by cell $(i,j)$. In this work, each cell $(i,j)$ is modeled by an equivalent circuit model sketched in Figure \ref{ECM}.}
\label{fig:Pack}
\end{figure*}
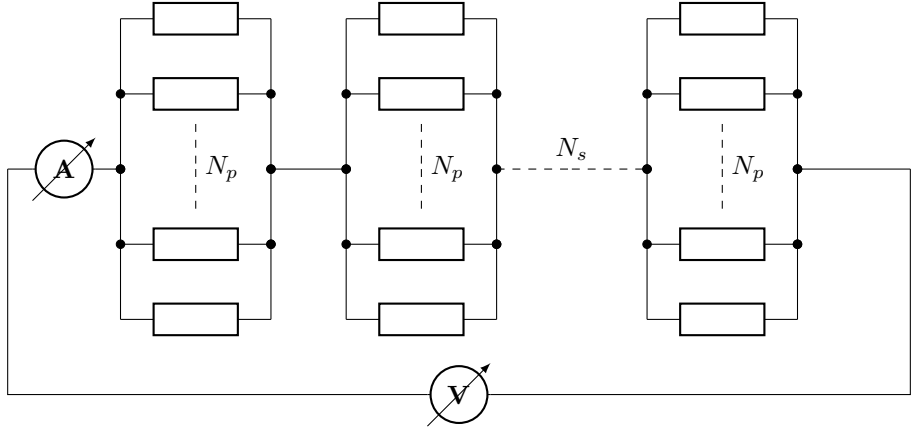

The battery pack configuration considered in this study is shown in Figure \ref{fig:Pack}. It is composed of $N_s$ battery modules connected in series, and each module contains $N_p$ battery cells connected in parallel. Such configuration represents a wide range of EV battery packs, e.g., Tesla Model S and Model 3. We assume that all cells are heterogeneous. Namely, they may be characterized by different model parameters (capacity, internal resistance etc.), different SOC levels, and different temperature distribution \cite{zhang2020interval}. The above-mentioned heterogeneity among cells can be caused by manufacturing, temperature variability, and battery degradation. Moreover, in order to reduce hardware sensing efforts, we further assume that only the pack-level voltage and the current of the string are available for measurement (see Figure \ref{fig:Pack}). This is referred to as the minimal sensing scenario, in which the voltage and current of each single cell in the pack are not explicitly under surveillance. It is noteworthy that reduced sensing significantly lowers the cost of battery design and assembly by possibly minimizing the required number of distributed sensors, but at the cost of diminishing the ability to monitor cell-level real-time conditions. This calls for a significant incentive to accurately estimate all cell-level SOCs under reduced sensing.

\begin{figure}[t]
	\centering
	\includegraphics[trim = 5.5mm 3.5mm 8.5mm 10mm, clip, width=0.7\linewidth]{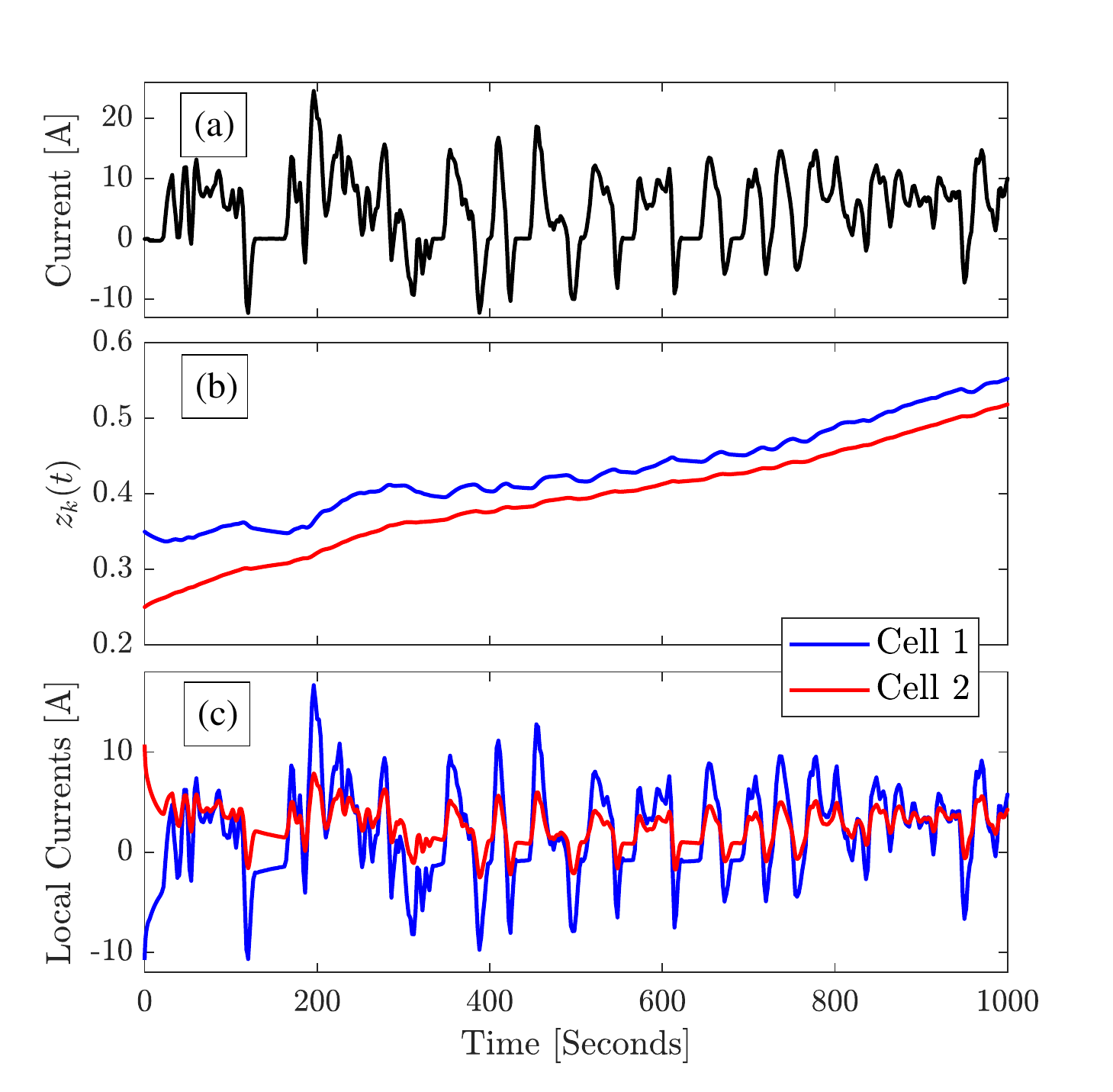}
    \caption{Simulation results of two cells in parallel using coupled electrical-thermal dynamics with temperature and SOC dependent electrical parameters. In (b)-(c), cells are initialized at different SOCs. The total current distributes unevenly due to both parameter and initialization heterogeneities.}
	\label{Motivation_Parallel}
\end{figure}

According to Kirchhoff's laws, the measured pack-level total voltage is equal to the summation of all module voltages, and the currents entering every parallel module are the same. However, since cells are heterogeneous, it is expected that the distributions of the total current to local branches in a module are imbalanced. We demonstrate this via an open-loop numerical simulation for two cells connected in parallel in Figure \ref{Motivation_Parallel}. This numerical study utilizes two Graphite/NMC type cells with 2.8 Ah nominal capacity. In this embodiment, the cells are parameterized distinctly but have identical SOC-OCV relationship. A transient electric vehicle-like duty charge/discharge cycle generated from the standardized Urban Dynamometer Driving Schedule (UDDS) is applied. Moreover, we initialize the cells at different SOCs. Specifically, the total applied current (summation of local cell currents) is plotted in Figure \ref{Motivation_Parallel}(a). It can be observed that even though the applied total current is small initially (around zero, see Figure \ref{Motivation_Parallel}(a)), Cell 1 draws a large negative current (around $-10$ A) and Cell 2 positions itself at a large positive current (around $+10$ A). 
This occurs because Cell 1 SOC is initialized higher (see Figure \ref{Motivation_Parallel}(b)). 
In such cases, single cells can violtae safety constraints (e.g. maximum current, see Figure \ref{Motivation_Parallel}(c)) while the pair of parallel cells do not, whereas the latter is the one that is commonly supervised. 
In the long run, the values for SOC follow a similar trend while seemingly approaching, but they never converge and cells accept different current rates, which promote different aging patterns and further increase cell-to-cell variations.
This behavior can be ascribed to the parallel connection that forces a natural voltage balance, but not SOC balance, between the cells.

\begin{figure}[t]
	\centering
	\includegraphics[trim = 5.5mm 3.5mm 8.5mm 9mm, clip, width=0.7\linewidth]{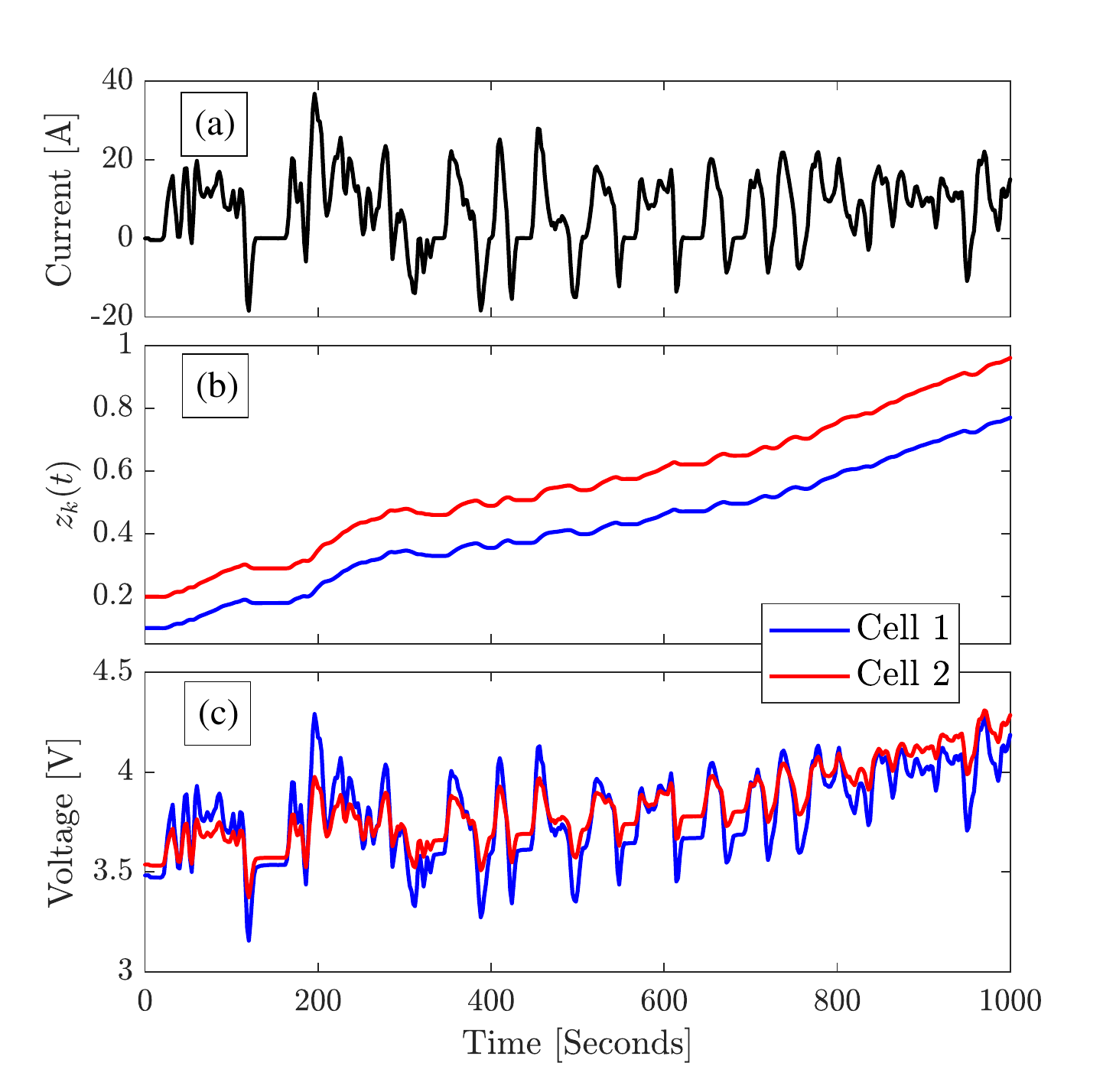}
    \caption{Simulation results of two cells in series using coupled electrical-thermal dynamics with temperature and SOC dependent electrical parameters. In (b)-(c), the initial cell SOCs are distinct. This discrepancy will persist because all cells accept the same input current.}
	\label{Motivation_Series}
\end{figure}

Furthermore, we encapsulate the case of the series arrangement of two heterogeneous cells, shown in Figure \ref{Motivation_Series}. This can be considered as a milder scenario compared to the parallel case in terms of local current behaviors, since every cell accepts identical and possibly measurable current. 
However, SOC discrepancy is worse and more persistent in time (see Figure \ref{Motivation_Series}(b)) than the parallel case, since the SOC values for the two cells will never synchronize -- a bias will always exist unless an external active action is taken, such as cell balancing. 

Under reduced/minimal sensing, the battery management systems do not monitor the local current and local voltage of each cell in a pack, which might translate in some cells operating outside their safe operating region whereas this cannot be seen from the pack-level information. Therefore, it will be of significant value to monitor local SOCs, currents, and voltages caused by cell heterogeneity in order to ensure safe battery pack operation and mitigate cell degradation.

\section{Battery Pack Model Formulation} \label{s:mod}

This section first reviews an equivalent circuit model (ECM) for a single battery cell, which is then electrically interconnected with other cell models to form a pack model. Although an ECM is employed in this work, the observer design and analysis can be readily generalized to other types of battery cell models, e.g., the reduced-order electrochemical models \cite{moura2016-electrolyte,zhang2019stress}.

\subsection{Single Cell Model}

\begin{figure}[t]
 \centering
\resizebox{0.6\linewidth}{!}{
\begin{circuitikz}[american, ]
\draw
(2,0) to[short, *-*] (3,0)
(0,0) to[short, -*] (0.5,0)
(-2.5,-2) to[short, -*] (3,-2)
(0.5,0) -- (0.5,-0.8) to[C,l_=$C_{i,j}$] (2,-0.8) -- (2,0)
(0.5, 0) to [open,v=$U_{i,j}$] (2,0)
(0.5,0) -- (0.5,0.8) to[R,l=$R_{i,j}$] (2,0.8) -- (2,0)
(-2.5,0) to [V, l_=$g({z_{i,j}})$] (-2.5,-2)
(-2.5,0) to[R, l^=$r_{i,j}$, i=$I_{i,j}$] (0,0);

\node at (3,-1) {$V_{i,j}$};
\node at (3,0) (s1) {};
\node at (3,-0.8) (s2) {};
\draw[<-](s1) -- node {} (s2);
\node at (3,-1.2) (s3) {};
\node at (3,-2) (s4) {};
\draw[->](s3) -- node {} (s4);
\end{circuitikz} 
}
\caption{The schematic of an equivalent circuit model, which is leveraged to model each cell in the pack given by Figure \ref{fig:Pack}.}
\label{ECM}
\end{figure}
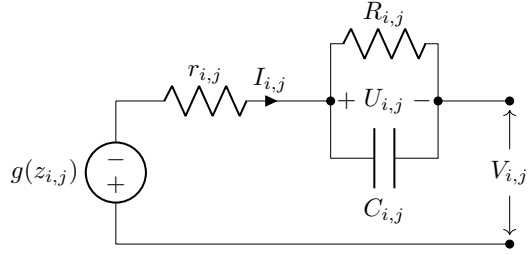

Consider the ECM for a single battery cell \cite{zhang2017remaining}, shown in \mbox{Figure \ref{ECM}}, represented by the following continuous-time state-space representation,
\begin{align}
    \label{e:ecm}
    \dot{s}_{i,j}(t) & = \overline{A}_{i,j} s_{i,j}(t) + \overline{B}_{i,j} I_{i,j}(t), \\
    \label{e:oecm}
    y_{i,j}(t) & = V_{i,j} = P(s_{i,j},I_{i,j}),
\end{align}
where $s_{i,j} = [z_{i,j} \quad U_{i,j}]^\top \in \mathbb{R}^2$ is the state vector for the $i$-th battery cell in the \mbox{$j$-th} parallel module, depicted in Figure \ref{ECM}, in which $z_{i,j}$ is cell SOC and $U_{i,j}$ represents the relaxation voltage for the parallel $R$-$C$ pair. In \eqref{e:ecm}-\eqref{e:oecm}, $I_{i,j}(t)$ denotes the current passing through the $i$-th battery cell in the \mbox{$j$-th} parallel module and $V_{i,j} \in \mathbb{R}$ is the cell terminal voltage. The state matrix $\overline{A}_{i,j} {\in \mathbb{R}^{2\times2}}$ and input matrix $\overline{B}_{i,j}{\in \mathbb{R}^{2\times 1}}$ are given by
\begin{equation}
\overline{A}_{i,j} = \begin{bmatrix}0&0\\0&-\frac{1}{R_{i,j}C_{i,j}}\end{bmatrix}, \quad
\ \overline{B}_{i,j} = \begin{bmatrix}\frac{1}{Q_{i,j}} \\ \frac{1}{C_{i,j}}\end{bmatrix},
\label{e:ecmmat}
\end{equation}
where $Q_{i,j}$ represents battery capacity, and $r_{i,j}$, $R_{i,j}$, $C_{i,j}$ are resistances and capacitance shown in Figure \ref{ECM}. The output equation \eqref{e:oecm} for the $(i,j)$-th cell provides the voltage response characterized by the function
\begin{equation}
    \label{e:ohecm}
    P(s_{i,j},I_{i,j}) = g(z_{i,j}) + U_{i,j} + r_{i,j}I_{i,j}.
\end{equation}
In \eqref{e:ohecm}, function $g(\cdot)$ denotes the open circuit voltage (OCV) that is a nonlinear function with respect to $z_{i,j}$. The cell voltage $V_{i,j}:[0,1]\times\mathbb{R}\times\mathbb{R}\rightarrow \mathbb{R}$ is the summation of the OCV, the relaxation voltage $U_{i,j}$, and voltage contributed by the ohmic resistance $r_{i,j}$.

\begin{rmk}
Although the single cells in the battery pack under consideration are modeled by ECM for its structural simplicity, the analysis and algorithm designs are applicable to other modeling framework, e.g., classes of electrochemical models.
\end{rmk}

\subsection{Kirchhoff's Laws}

For a module of $N_p$ cells in parallel (i.e., one of the parallel modules in Figure \ref{fig:Pack}), Kirchhoff's voltage law indicates that a parallel connection constraints terminal voltages to the same value for all cells, and Kirchhoff's current law indicates that the total current is equivalent to the summation of all branch currents. Mathematically, the following nonlinear algebraic constraints, according to Kirchhoff's voltage law, need to be enforced,
\begin{align}
     \label{volt_constraint_comp}
    P(s_{k,j},I_{k,j}) = P(s_{\ell,j},I_{\ell,j}), \quad \forall k,\ell \in \{1,\cdots,N_p\}, \; k \neq \ell, \; j\in\{1,\cdots,N_s\},
\end{align}
which can be equivalently expanded as
\begin{align}
    \label{volt_constraint}
    g(z_{k,j}) + {} & U_{k,j} + r_{k,j} I_{k,j} =  g(z_{\ell,j}) + U_{\ell,j} + r_{\ell,j} I_{\ell,j}, \nonumber \\
    & \forall k,\ell \in \{1,\cdots,N_p\}, \; k \neq \ell, \; j\in\{1,\cdots,N_s\}.
\end{align}
Similarly, Kirchhoff's current law poses the following linear algebraic constraint with respect to cell local currents,
\begin{align}
    \label{cur_constraint}
    \sum_{i = 1}^{N_p} I_{i,j}(t) = I(t), \quad \forall j\in\{1,\cdots,N_s\},
\end{align}
where $I(t)$ is the measurable pack-level total current. It is worth highlighting that, for every parallel module $j$, equation (\ref{volt_constraint}) imposes $(N_p-1)$ nonlinear algebraic constraints with respect to local SOCs and local currents, whereas equation (\ref{cur_constraint}) imposes 1 additional algebraic constraint with respect to local currents. 

\subsection{Parallel Module Model}

When only the total current is measured, the local currents of cells are unknown since they are unevenly distributed across cells due to cell heterogeneity. Hence, a system of differential-algebraic equations must be solved such that the algebraic equations (\ref{volt_constraint}) and (\ref{cur_constraint}) are fulfilled at all times $t > 0$. Such methodology is realized by augmenting the local currents (algebraic states) to the differential states (local SOCs) to form an aggregated state vector for a nonlinear DAE system \cite{duan2010analysis}, which takes the form
\begin{align}
    \label{parallel_dyn}
    E_j \dot{x}_j(t) = {} & A_j x_j(t) + \phi(x_j(t),I(t)),
\end{align}
where $x_j = [s_j^\top \quad u_j^\top]^\top \in \mathbb{R}^{3N_p}$, $\forall j \in \{1,\cdots,N_s\}$, with
\begin{align}
    \label{s_j}
    s_j & = \begin{bmatrix}s_{1,j}^\top & s_{2,j}^\top & \cdots & s_{N_p,j}^\top \end{bmatrix}^\top \in \mathbb{R}^{2N_p}, \\
    \label{u_j}
    u_j & = \begin{bmatrix}I_{1,j} & I_{2,j} & \cdots & I_{N_p,j}\end{bmatrix}^\top \in \mathbb{R}^{N_p}.
\end{align}
Equation (\ref{parallel_dyn}) encodes both the dynamical equations \eqref{e:ecm} and algebraic constraints \eqref{volt_constraint}-\eqref{cur_constraint} for the $j$-th parallel module, and the matrix $E_j$ is a singular matrix of the form
\begin{equation}
    \label{E_mat}
    E_j = \begin{bmatrix}\mathbf{I}_{2N_p \times 2N_p} & \mathbf{0}_{2N_p \times N_p} \\ \mathbf{0}_{N_p \times 2N_p} & \mathbf{0}_{N_p \times N_p}\end{bmatrix} \in \mathbb{R}^{3N_p \times 3N_p}.
\end{equation}
Matrix $A_j$ accounts for the linear part of the system equations with
\begin{equation}
    A_j = \begin{bmatrix} A_{11,j} & A_{12,j} \\ A_{21,j} & A_{22,j} \end{bmatrix} \in \mathbb{R}^{3N_p \times 3N_p}, \label{e:Amatrix}
\end{equation}
where
\begin{align}
    A_{11,j} & = \mathrm{diag}(\overline{A}_{1,j}, \; \overline{A}_{2,j}, \; \cdots, \; \overline{A}_{N_p,j}), \nonumber \\ 
    A_{12,j} & = \mathrm{diag}(\overline{B}_{1,j}, \; \overline{B}_{2,j}, \; \cdots, \; \overline{B}_{N_p,j}), \nonumber \\
    A_{21,j} & = \begin{bmatrix}
    0 & 1 & S & 0 & \cdots & 0 \\
    0 & 1 & 0 & S & \cdots & 0 \\
    \vdots & \vdots & \vdots & \vdots & \ddots & \vdots \\
    0 & 1 & 0 & 0 & \cdots & S \\
    0 & 0 & 0 & 0 & \cdots & 0 \\
    \end{bmatrix} \in \mathbb{R}^{N_p \times 2N_p}, \;\text{where }
    S = \begin{bmatrix} 0 & -1 \end{bmatrix}, \nonumber 
\end{align}
\begin{align}
    A_{22,j} = \begin{bmatrix}
    r_{1,j} & -r_{2,j} & 0 & \cdots & 0 \\
    r_{1,j} & 0 & -r_{3,j} & \cdots & 0 \\
    \vdots & \vdots & \vdots & \ddots & \vdots \\
    r_{1,j} & 0 & 0 & \cdots & -r_{N_p,j} \\
    1 & 1 & 1 & \cdots & 1 \\
    \end{bmatrix} \in \mathbb{R}^{N_p \times N_p}.
    \label{e:A22mat}
\end{align}
Notice that matrix $A_{22,j}$ is full rank \cite{drummond2020resolving}, i.e. the linear part of the DAE model \eqref{parallel_dyn} is regular and impulsive free \cite{duan2010analysis}. Function $\phi(x_j,I)$ in \eqref{parallel_dyn} constitutes the nonlinear portion in the system equations from the voltage algebraic constraints (\ref{volt_constraint}),
\begin{equation}
    \label{phifunc}
    \phi(x_j,I) = \begin{bmatrix}
    \phi_s \\
    \phi_u
    \end{bmatrix} = 
    \begin{bmatrix}
    \mathbf{0}_{2N_p \times 1} \\
    g(z_{1,j}) - g(z_{2,j}) \\ 
    g(z_{1,j}) - g(z_{3,j}) \\
    \vdots \\
    g(z_{1,j}) - g(z_{N_p,j}) \\
    -I(t)
    \end{bmatrix} \in \mathbb{R}^{3N_p},
\end{equation}
where $\phi_s$ represents the nonlinearities in the dynamical equations and corresponds to row 1 through row $2N_p$ of $\phi$. Evidently we have $\phi_s = \mathbf{0}_{2N_p \times 1}$ due to the linear nature of system \eqref{e:ecm}. $\phi_u \in \mathbb{R}^{N_p}$ encodes the nonlinearities and input appearing in the algebraic constraints, and it corresponds to row $(2N_p+1)$ through row $3N_p$ of $\phi$.

\subsection{Pack Model}

For a string of parallel modules wired in series \mbox{(Figure \ref{fig:Pack})}, there are in total $N = N_p \times N_s$ cells. The measured pack-level voltage is equivalent to the summation of all parallel module voltages, and every parallel module shares the same current, according to Kirchhoff's laws. Let us denote $n = 2N$ and
\begin{align}
    x = \begin{bmatrix} x_{1}^\top & x_{2}^\top & \cdots & x_{N_s}^\top \end{bmatrix}^\top \in \mathbb{R}^{n},
\end{align}
where respective $x_j$ (the state vector for parallel module $j$) was previously introduced in \eqref{parallel_dyn}-\eqref{u_j}. $x$ is essentially the state vector that aggregates all differential (local SOCs) and algebraic (local currents) states of all cells in the pack. Based on the parallel module model in \eqref{parallel_dyn}, a string of $N_s$ parallel modules can be expressed in the compact form
\begin{align}
    \label{pack_dyn}
    E\dot{x}(t) = {} & Ax(t) + \Phi(x(t),I(t)), \\
    \label{pack_out}
    y(t) = {} & H(x(t)) = Cx(t) + h(x(t)),
\end{align}
in which $y(t) = V(t)$ is the measured pack voltage, and
\begin{align}
    E = {} & \mathrm{diag}(E_{1}, \; E_{2}, \; \cdots, \; E_{N_s}), \nonumber \\
    A = {} & \mathrm{diag}(A_{1}, \; A_{2}, \; \cdots, \; A_{N_s}), \nonumber \\
    \Phi(x,I) = {} & \begin{bmatrix} \phi(x_1,I)^\top & \phi(x_2,I)^\top & \cdots & \phi(x_{N_s},I)^\top \end{bmatrix}^\top.
\end{align}
where $E_j$, $A_j$, and $\phi(x_j,I)$ have been introduced in \eqref{e:Amatrix} , \eqref{E_mat}, and \eqref{phifunc}, respectively. In \eqref{pack_out}, the pack voltage expression has been partitioned into a linear part $Cx(t)$ and a nonlinear part $h(x(t))$, which will be detailed later.
Note that in the subsequent sections, we will slightly abuse the notations for the ease of analysis and presentation. Specifically, in \eqref{pack_dyn}, the state vector $x$ is composed of $x_j$, $j\in\{1,2,\cdots,N_s\}$, concatenated according to to module numbers. Nonetheless, from now on, we alter the sequence in $x$ such that the differential state vector $s = [s_1^\top \quad s_2^\top \quad \cdots \quad s_{N_s}^\top]^\top$ comes before the algebraic state vector $u = [u_1^\top \quad u_2^\top \quad \cdots \quad u_{N_s}^\top]^\top$, i.e.,
\begin{equation}
    \label{xsu}
    x = \begin{bmatrix} s^\top & u^\top \end{bmatrix}^\top.
\end{equation}
The system matrices $E$, $A$, and $\Phi$ are adjusted accordingly. In that event, we also let the partition of the function $\Phi$ to be $\Phi(x,I) = [\Phi_s(x) \quad \Phi_u(x,I)]^\top$, where vector $\Phi_s$ accounts for the nonlinearities in the differential equations whereas $\Phi_u$ lumps the nonlinearities in the algebraic constraints. Note, \mbox{$\Phi_s \equiv \mathbf{0}_{N}$} since battery dynamics described by the ECM \eqref{ECM} is linear. However, we maintain the appearances of $\Phi_s$ in the forthcoming analysis to make the framework generalizable to other nonlinear battery dynamics.

Special care needs to be taken for the output function $H(x(t))$ in (\ref{pack_out}). $H(x(t))$ represents the summation of all parallel module voltages, while each parallel module voltage can be mathematically expressed in $N_p$ different ways. Namely, each parallel module voltage is equivalent to any cell voltage in that module. Thus, in order to increase the information contained in the pack system output to maximize system observability, $H(x(t))$ must include all $(N_p)^{N_s}$ possible combinations, although the numerical values of them would all equal to the measured pack voltage signal:
\begin{align}
\label{pack_out_2}
& H(x(t)) = \begin{bmatrix} V(t) \\ V(t) \\ \vdots \\ V(t)\end{bmatrix}^\top = \begin{bmatrix} V_{1,1} + V_{1,2} + \cdots + V_{1,N_s-1} + V_{1,N_s} \\ V_{1,1} + V_{1,2} + \cdots + V_{1,N_s-1} + V_{2,N_s} \\ \vdots \\ V_{N_p,1} + V_{N_p,2} + \cdots + V_{N_p,N_s-1} + V_{N_p,N_s} \end{bmatrix}^\top,
\end{align}
where each $V_{i,j}$ is given by \eqref{e:oecm}.

\begin{rmk}
In fact, not all $(N_p)^{N_s}$ combinations are unique. That is, only a subset of these combinations are linearly independent of others. As an example, consider a simple case with $N_p = 2$ and $N_s = 2$. In this case, 4 possible voltage representations can be generated, namely
\begin{align}
    H(x(t)) = \begin{bmatrix} y_1 \\ y_2 \\ y_3 \\ y_4 \end{bmatrix} = \begin{bmatrix} (V_{1,1} + V_{1,2}) \\ (V_{1,1} + V_{2,2}) \\ (V_{2,1} + V_{1,2}) \\ (V_{2,1} + V_{2,2}) \end{bmatrix}.
\end{align}
Nevertheless, it is easily observed that $y_4 = y_2 + y_3 - y_1$, making $y_4$ a redundant output entry. Ultimately, the formulation of $H(x(t))$ in \eqref{pack_out_2} should be carefully calibrated based upon the structures of the pack to remove redundant entries, which effectively reduces the mathematical complexity in the observability analysis in the upcoming Section \ref{s:obs_ana}.
\end{rmk}

\begin{rmk}
\label{DAE_RED}
In Kirchhoff's laws \eqref{volt_constraint}-\eqref{cur_constraint}, although these algebraic constraints are nonlinear in the differential states $z_{i,j}$ (individual cell SOCs), they are indeed linear in the algebraic states $I_{i,j}$ (individual cell currents). Consequently, in a parallel module, one can solve for the algebraic states as a closed form of differential states and substitute this expression into the system dynamics to obtain a reduced-order ODE system, which is ultimately independent of the algebraic states. This strategy was carried out in \cite{drummond2020resolving,fan2020simplified,hofmann2018dynamics}. In spite of the fact that this model reduction from DAEs to ODEs is useful for simulation studies to understand imbalanced current distributions in parallel and parallel-series configurations, the reduced-order ODEs are mathematically not the exact representation of the original DAEs \cite{petzold1982differential}, primarily due to the following reasons: (i) numerical
methods that are commonly used for solving systems of ODEs do not trivially apply to DAEs; (ii) the solution to the reduced-order ODEs does not always satisfy the algebraic constraints; and (iii) when a state estimator is designed for the reduced-order ODEs, the algebraic state will only be estimated in an open-loop fashion, whereas the differential states will most likely be estimated in a feedback (closed-loop) manner. Ultimately, due to these restrictiveness and limitations, we never perform model reduction in order to retain the physical significance of the differential-algebraic nature and of the phenomena that they represent. All analysis and observer designs are based upon the high-fidelity DAE system \eqref{pack_dyn}-\eqref{pack_out}. This, together with the minimal sensing setup, substantially separates our method from the existing battery pack state estimation works in the literature, which tend to lump the entire pack as a single virtual cell whose dynamics is described by a reduced-order ODE that significantly loses the tractability of individual cells or conduct the measurement of few specific cells inside a pack.
\end{rmk}

\begin{rmk}
In practice, the parallel modules in Figure \ref{fig:Pack} are wired with the same number of cells, but the proposed pack modeling framework and the subsequent state observability analysis in Section \ref{s:obs_ana} can be readily generalized to the case when the parallel modules have different number of cells under extreme circumstances.
\end{rmk}

The battery pack model introduced above will be leveraged in the analysis and designs in the subsequent sections.

\section{Observability Analysis} \label{s:obs_ana}

Informally, observability analysis refers to the study of the conditions under which it is possible to uniquely determine the states of a dynamical system from measurements of its input and output \cite{Vidal_Obs}. In other words, a system, e.g., \eqref{pack_dyn}-\eqref{pack_out}, is said to be \textit{completely observable} on an interval $(t_0,t_1)$ if the initial state can be uniquely determined from knowledge of the output $y(t)$ and input $I(t)$ over $(t_0,t_1)$ \cite{silverman1967controllability}. Specifically for our case, if the battery pack system is unobservable, this means that it is impossible to infer the SOC of at least some cells from the pack-level voltage and current data. In this section, we mathematically analyze the conditions for the observability of individual cell SOC in a battery pack under the minimal sensing scenario (Figure \ref{fig:Pack}), in which only pack-level voltage and current are measured. For mathematical tractability, the observability conditions are derived from the pack configuration shown in Figure \ref{fig:Pack} with $N_s = 2$ and $N_p = 2$ ($N = 4$ cells in total), which is conveniently referred to as 2P2S. We study the observability using two techniques, namely (i) observability via linearization of the pack model \eqref{pack_dyn}-\eqref{pack_out}, and (ii) nonlinear DAE smooth observability inherited from DAE solvability. 

Previously in the literature, observability analysis has been carried out for cells connected in series \cite{Lin-2015} and in parallel \cite{ZhangDescriptor} under reduced sensing scenarios. The former assumes total voltage measurement for a string of heterogeneous cells, and the latter faces unknown imbalanced current distributions in a parallel module. Both studies require non-flatness of high-order gradient of cell's OCV function.
The observability matrices for the series case \cite{Lin-2015} are derived from a conventional ODE setting, differing from that of a parallel (and series-parallel considered in this work) arrangement. Namely, in the series arrangement, each cell's parameters/states appear in a column of observability matrix. See, for instance, equation (10) in \cite{Lin-2015}. 
This is not the case for a battery pack with parallel topology, where parameters/states of the cells are scattered all over the different entries in the observability conditions (See Section IV in \cite{ZhangDescriptor}). Therefore, parameters/states of one cell influence the observability of the neighboring cells in a pack containing parallel modules. 
Importantly, in light of Remark \ref{DAE_RED}, the observability analysis in this section will be conducted directly on the pack DAE system, rather than a reduced-order ODE system, to examine if both differential and algebraic states are observable.

\begin{rmk}
The dynamics for the cell relaxation voltage $U_{i,j}$ in \eqref{e:ecm} exponentially decays to an equilibrium position. This state is generally observable (at least detectable) from current-voltage measurements. In the subsequent observability analysis and observer designs, the effect of relaxation voltage is ignored to simplify computations.
\end{rmk}

\subsection{Observability via Model Linearization}
\label{sec:Linear_Obs}

To study the observability of any general form of a nonlinear system, one of the most convenient strategies is to linearize the system around an equilibrium point and derive the local observability conditions for the linearized system. If the linearized system is observable at a given equilibrium point, then the nonlinear system is locally observable at that position. This method is practically easy to implement but the results are only sufficient. That is, the observability conditions arising from linearizing the nonlinear system can be conservative, and nothing can be concluded for the original nonlinear system if the linearized system is \emph{not} observable. Thus, under the unobservable condition from the linearized system, less conservative observability notions need to be explored. In this section, we first study if local observability exists from the linearization of the nonlinear pack DAE system \eqref{pack_dyn}-\eqref{pack_out}.

The DAE system \eqref{pack_dyn}-\eqref{pack_out} with $N_s = 2$ and $N_p = 2$ is employed to describe the dynamics for 2P2S with $x  = [x_{1} \;\; x_{2} \;\; x_{3} \;\; x_{4} \;\; x_{5} \;\; x_{6} \;\; x_{7} \;\; x_{8}]^\top = [z_{1,1} \;\; z_{2,1} \;\; z_{1,2} \;\; z_{2,2} \;\; I_{1,1} \;\; I_{2,1} \;\; I_{1,2} \;\; I_{2,2}]^\top \in \mathbb{R}^8$, and 
\begin{align}
    A = & \begin{bmatrix} 0 & 0 & 0 & 0 & \frac{1}{Q_{1,1}} & 0 & 0 & 0 \\ 0 & 0 & 0 & 0 & 0 & \frac{1}{Q_{2,1}} & 0 & 0 \\ 0 & 0 & 0 & 0 & 0 & 0 & \frac{1}{Q_{1,2}} & 0 \\  0 & 0 & 0 & 0 & 0 & 0 & 0 &\frac{1}{Q_{2,2}} \\ 0 & 0 & 0 & 0 & 1 & 1 & 0 & 0 \\ 0 & 0 & 0 & 0 & 0 & 0 & 1 & 1 \\ 0 & 0 & 0 & 0 & r_{1,1} & -r_{2,1} & 0 & 0 \\  0 & 0 & 0 & 0 & 0 & 0 & r_{1,2} & -r_{2,2}\end{bmatrix}, \; \Phi(x,I) = \begin{bmatrix} 0 \\ 0 \\ 0 \\ 0 \\ -I \\ -I \\ g(x_1) - g(x_2) \\ g(x_3) - g(x_4)
    \end{bmatrix},\nonumber \\
     E = & \begin{bmatrix} 1 & 0 & 0 & 0 & 0 & 0 & 0 & 0 \\ 0 & 1 & 0 & 0 & 0 & 0 & 0 & 0 \\ 0 & 0 & 1 & 0 & 0 & 0 & 0 & 0 \\  0 & 0 & 0 & 1 & 0 & 0 & 0 & 0 \\ 0 & 0 & 0 & 0 & 0 & 0 & 0 & 0 \\ 0 & 0 & 0 & 0 & 0 & 0 & 0 & 0 \\ 0 & 0 & 0 & 0 & 0 & 0 & 0 & 0 \\  0 & 0 & 0 & 0 & 0 & 0 & 0 & 0\end{bmatrix}, \;
    \label{H}
    H(x) = \begin{bmatrix} g(x_1) + r_{1,1}x_5 + g(x_3) + r_{1,2}x_7 \\ g(x_1) + r_{1,1}x_5 + g(x_4) + r_{2,2}x_8 \\ g(x_2) + r_{2,1}x_6 + g(x_3) + r_{1,2}x_7
    \end{bmatrix}.
\end{align}
To study the observability of the pack model \eqref{pack_dyn}-\eqref{pack_out} with \eqref{H}, we linearize the system around an equilibrium point $x = \overline{x}$ and check the rank of the observability matrix of the linearized system. The linearized model takes the form
\begin{align}
    E \dot{x}(t) = {} & T x(t) + B I(t), \label{e:linstate} \\
    y(t) = {} & C x(t), \label{e:linout}
\end{align}
where the state matrix $T\in\mathbb{R}^{8 \times 8}$ and output matrix $C\in\mathbb{R}^{3 \times 8}$ are given by
\begin{equation}
T = \left. A + \frac{\mathrm{d} \Phi}{\mathrm{d} x}(x)\right\vert_{x=\overline{x}}, \quad
C = \left. \frac{\mathrm{d} H}{\mathrm{d} x}(x)\right\vert_{x=\overline{x}},
\end{equation}
with matrix $A$ and $H$ given in \eqref{H}. Matrices $T$ and $C$ take the form
\begin{align}
T & = \begin{bmatrix} 0 & 0 & 0 & 0 & \frac{1}{Q_{1,1}} & 0 & 0 & 0 \\ 0 & 0 & 0 & 0 & 0 & \frac{1}{Q_{2,1}} & 0 & 0 \\ 0 & 0 & 0 & 0 & 0 & 0 & \frac{1}{Q_{1,2}} & 0 \\  0 & 0 & 0 & 0 & 0 & 0 & 0 &\frac{1}{Q_{2,2}} \\ 0 & 0 & 0 & 0 & 1 & 1 & 0 & 0 \\ 0 & 0 & 0 & 0 & 0 & 0 & 1 & 1 \\ g'(x_1) & -g'(x_2) & 0 & 0 & r_{1,1} & -r_{2,1} & 0 & 0 \\ 0 & 0 & g'(x_3) & -g'(x_4) & 0 & 0 & r_{1,2} & -r_{2,2}\end{bmatrix}, \nonumber \\
C & = \begin{bmatrix}
    g'(x_1) & 0 & g'(x_3) & 0 & r_{1,1} & 0 & r_{1,2} & 0 \\
    g'(x_1) & 0 & 0 & g'(x_4) & r_{1,1} & 0 & 0 & r_{2,2} \\
    0 & g'(x_2) & g'(x_3) & 0 & 0 & r_{2,1} & r_{1,2} & 0 \\
    \end{bmatrix}.
\end{align}
Let us now introduce the definition of Complete Observability (C-Observability) for the linearized DAE system \eqref{e:linstate}-\eqref{e:linout}, which represents a linearized battery pack model for the topology in Figure \ref{fig:Pack}.

\begin{thm} [Complete Observability \cite{duan2010analysis}]
\label{thm_lin_obs}
The regular linear descriptor system \eqref{e:linstate}-\eqref{e:linout} is \textit{Complete Observable} if and only if the following two conditions hold:
\begin{enumerate}[(C1)]
    \item $\mathrm{rank}([ E^\top, C^\top ]^\top) = 2N$;
    \item $\mathrm{rank}([(\lambda E - T)^\top, C^\top ]^\top) = 2N, \ \forall \lambda \in \mathbb{C}$.
\end{enumerate}
\end{thm}

It follows from \cite{duan2010analysis} that there exists a standard decomposition, such that the descriptor linear system \eqref{e:linstate}-\eqref{e:linout} is transformed into a slow subsystem (an ODE system) and a fast subsystem (a DAE subsystem). The system \eqref{e:linstate}-\eqref{e:linout} is C-Observable if and only if both its slow and fast subsystems are observable. See Section 4 of \cite{duan2010analysis} for more precise definitions.
Specifically, Condition (C1) concerns observability of the fast (algebraic) subsystem while (C2) involves the slow (differential) subsystem observability.
Focusing first on condition (C1), it can be verified that the expression in (C1) is rank deficient. Intuitively, the singular matrix $E$ offers rank $N = 4$ with its non-zero columns. Nonetheless, matrix $C$ does not have enough rows (or measurements) to fulfill the missing rank. This means that the pack voltage measurement is potentially insufficient to uniquely determine the branch current of every cell.
Looking at condition (C2), which needs to be verified for all $\lambda \in \mathbb{C}$. Fortunately, for those $\lambda$'s that are not one of the generalized eigenvalues of the pair $(E,T)$, condition (C2) automatically validates.
Thus, the verification of condition (C2) requires the numerical computation of the generalized eigenvalues of the pair $(E,T)$, by solving the characteristic equation $\det(\lambda E - T) = 0$, and $\lambda = 0$ is guaranteed to be one of the solutions to this equation. When $\lambda = 0$, condition (C2) boils down to $\mathrm{rank}( [ - T^\top, C^\top ]^\top ) = 2N$, which is not true. Consequently, neither conditions in Theorem \ref{thm_lin_obs} can be verified. Hence, we cannot draw any conclusions towards the individual cell SOC and current observability under minimal sensing by checking the observability conditions of the linearized DAE system. This incentivizes a more sophisticated (less conservative) strategy to derive the observability conditions of the original nonlinear DAE system \eqref{pack_dyn}-\eqref{pack_out}.

\subsection{DAE Solvability}

Before introducing a less conservative notion of observability, let us first present the concept  of solvability of a DAE system. Consider a general nonlinear implicit DAE of the form
\begin{align}
    \label{DAE_dyn}
    F(t,x(t),\dot{x}(t)) = {} & u(t), \\
    \label{DAE_out}
    y(t) = {} & H(x,t),
\end{align}
with $x \in \mathbb{R}^{n}$, $F(\cdot,\cdot,\cdot) \in \mathbb{R}^{n}$, and $F_{\dot{x}} = \partial F / \partial \dot{x}$ identically singular \cite{brenan1995numerical}. Note that the pack model \eqref{pack_dyn}-\eqref{pack_out} can be easily reformulated into this structure. We also assume that $H$ (or equivalently OCV) is sufficiently smooth so that the high-order derivatives with respect to states are continuous. There have been considerable amount of research on computing a solution for DAE \eqref{DAE_dyn}-\eqref{DAE_out}, which is a system of equations in the $(2n + 1)$-dimensional variable $(t, x, \dot{x})$. Let us now first provide the formal definition of \textit{solvability} of DAE \eqref{DAE_dyn} \cite{campbell1995solvability,campbell1993least}.

\begin{definition}[Solvability \cite{campbell1995solvability}]
\label{solve_def}
DAE \eqref{DAE_dyn} is \textit{solvable} in an open set $\Omega \subseteq \mathbb{R}^{2n+1}$ if the graphs $(t, x, \dot{x})$ of the solutions form a smooth manifold in $\Omega$ called the \textit{solution manifold} and solutions are uniquely determined by their value $x_0$ at any $t_0$ such that $(t_0, x_0, \dot{x}_0) \in \Omega$.
\end{definition}

In general, the solution $x$ of DAE \eqref{DAE_dyn} is dependent on derivatives of $F$. If \eqref{DAE_dyn} is differentiated $\gamma$ times with respect to $t$, we get $(\gamma+1)n$ equations \cite{campbell1995solvability}:
\begin{align}
    \label{F_gam}
    \mathcal{F}_\gamma(t,x,\dot{x},w) = \begin{bmatrix} F(t,x,\dot{x}) \\ F_t(t,x,\dot{x}) + F_x(t,x,\dot{x})\dot{x} + F_{\dot{x}}(t,x,\dot{x})x^{(2)} \\ \vdots \\ \dfrac{{\rm d}^\gamma}{{\rm d}t^\gamma}F(t,x,\dot{x}) \end{bmatrix} = \mathbf{u},
\end{align}
where $w$ denotes the high-order derivatives of $x$, i.e., $w = [x^{(2)} \;\; x^{(3)} \;\; \cdots \;\; x^{(\gamma+1)}]$,
and $\mathbf{u}$ is a column vector with $\mathbf{u} = [u \;\; \dot{u} \;\; \cdots \;\; u^{(\gamma)}]^\top$.

While the definition for DAE solvability in Definition \ref{solve_def} is obscure, the assumptions below are verifiable sufficient
conditions for DAE solvability:
\begin{enumerate}[(S1)]
\item $\mathcal{F}_{\gamma}$ is sufficiently differentiable in its arguments.
\item $\mathbf{G} \equiv \mathcal{F}_{\gamma} = 0$ is consistent as an algebraic equation.
\item $[\mathbf{G}_{\dot{x}} \quad \mathbf{G}_{w}]$ is 1-full and has constant rank.
\item $[\mathbf{G}_x \quad \mathbf{G}_{\dot{x}} \quad \mathbf{G}_{w}]$ has full row rank independent of $(t,x,\dot{x},w)$.
\end{enumerate}
The minimum $\gamma$ for which conditions (S1)-(S4) hold is called the \textit{uniform differentiation index} \cite{campbell1995index}. In addition, $[\mathbf{G}_{\dot{x}} \quad \mathbf{G}_{w}]$ is 1-full with respect to $\dot{x}$ if the first $N$ columns are linearly independent, and linearly independent of the remaining columns.

The definitions and assumptions in this subsection will be utilized to establish the smooth observability in the forthcoming sections.

\subsection{Smooth Observability}
\label{smooth_obs}

Observability analysis using the linearized system failed as has been demonstrated in Section \ref{sec:Linear_Obs}. This motivates us to study a stronger notion of observability. To elucidate if less conservative observability conditions exist, we analyze the local observability of the nonlinear battery pack DAE system \eqref{pack_dyn}-\eqref{pack_out} by introducing the concept of smooth observability.

\begin{definition} [Smooth Observability \cite{campbell1991observability}]
\label{def:SO}
The nonlinear DAE system \eqref{DAE_dyn}-\eqref{DAE_out} is \textit{smoothly observable} (of order $(\delta,\gamma)$) on the interval $\mathcal{K}$ if there exists smooth functions $\Delta_l(t)$ and $\Gamma_l(t)$ on $\mathcal{K}$ such that
\begin{equation}
    \label{x_sm_obs}
    x(t) = \sum_{l = 0}^{\delta} \Delta_l(t)y^{(l)}(t) + \sum_{l = 0}^{\gamma} \Gamma_l(t)u^{(l)}(t).
\end{equation}
\end{definition}
Note that {\it smooth} is defined to mean infinitely differentiable, and the infinite differentiability is only used to make sure the observability conditions are necessary and sufficient \cite{campbell1991observability}.
In order to verify smooth observability, \eqref{x_sm_obs} is used to determine the least $\delta$ and $\gamma$ such that the solution of the DAE model $x(t)$ can be represented by a weighted sum of the time derivatives of input $u(t)$ and output $y(t)$. The essence of smooth observability is that, if $C$ in \eqref{e:linout} is not full column rank, then additional information to determine $x$ is obtained by differentiating the measurements.

\begin{rmk}
It should be emphasized that there exist various forms of observability other than the previously noted complete and smooth observability in the literature \cite{silverman1967controllability}. For instance, \textit{total observability} refers to complete observability on every sub-interval of $(t_0, t_1)$ and \textit{uniform observability} stands for smooth observability of order $(n-1, n-1)$. From Definition \ref{def:SO}, it is evident that smooth observability, which is selected in this work for analyzing battery pack system observability, is a stronger type of observability than total and complete observability. For deeper discussions on smooth observability, interested readers may refer to Section I of \cite{campbell1991observability}.
\end{rmk}

Differentiating the output expression \eqref{DAE_out} $\delta$ times with respect to $t$ yields
\begin{align}
    \label{H_eta}
    \mathbf{H} \equiv \mathcal{H}_\delta(t,x,\dot{x},w) = \begin{bmatrix} H(x) \\ H_x(x)\dot{x} \\ \vdots \\ \dfrac{{\rm d}^\delta}{{\rm d}t^\delta}H(x) \end{bmatrix} = \mathbf{y},
\end{align}
where $\mathbf{y} = [y \quad \dot{y} \quad \cdots \quad y^{(\delta)}]^\top$. Then we can express the combination of equation \eqref{F_gam} with equation \eqref{H_eta} as
\begin{equation}
    \mathcal{O}(t,x,\dot{x},w) = \begin{bmatrix} \mathbf{u} \\ \mathbf{y} \end{bmatrix}.
\end{equation}
Hence, the Jacobian matrix of $\mathcal{O}(t,x,\dot{x},w)$ with respect to $(x,\dot{x},w)$ is then given by
\begin{equation}
    \label{jacob}
    J_{\mathcal{O}} = \begin{bmatrix} \mathbf{G}_x & \mathbf{G}_{\dot{x}} & \mathbf{G}_w \\ \mathbf{H}_x & \mathbf{H}_{\dot{x}} & \mathbf{H}_w \end{bmatrix}.
\end{equation}
Now we are positioned to formally introduce the sufficient condition for smooth observability for the nonlinear DAE system \eqref{DAE_dyn}-\eqref{DAE_out}.

\begin{thm}[Smooth Observability Verifiable Conditions \cite{terrell1997observability}] 
\label{SO}
Suppose system \eqref{DAE_dyn} satisfies condtions (S1)-(S4) in a neighborhood $U$ of a consistent point $(t_0,x_0,\dot{x}_0,w_0)$. Additionally, suppose the Jacobian matrix $J_\mathcal{O}$ given in \eqref{jacob} validates
\begin{enumerate}[(O1)]
\item ${\rm rank}(J_{\mathcal{O}}) = n + {\rm rank}\left(\begin{bmatrix} \mathbf{G}_{\dot{x}} & \mathbf{G}_w \\ \mathbf{H}_{\dot{x}} & \mathbf{H}_w \end{bmatrix}\right)$, for $(t,x,\dot{x},w)\in U$
\item $J_{\mathcal{O}}$ has constant rank on $U$
\end{enumerate}
Then system \eqref{DAE_dyn}-\eqref{DAE_out} is smoothly observable on $U$.
\end{thm}

\begin{rmk}
The smooth observability conditions in Theorem \ref{SO} is local in a neighborhood of $U$, and the Jacobian matrix $J_\mathcal{O}$ depends on state $x$, OCV function $g$, and individual cell model parameters.
\end{rmk}

Utilizing Theorem \ref{SO}, we analyze the local smooth observability of the nonlinear battery pack DAE system \eqref{DAE_dyn}-\eqref{DAE_out} for 2P2S, in which
\begin{align}
    F(t,x(t),\dot{x}(t)) = \begin{bmatrix} \dot{x}_1 - x_5/Q_{1,1} \\ \dot{x}_2 - x_6/Q_{2,1} \\ \dot{x}_3 - x_7/Q_{1,2} \\ \dot{x}_4 - x_8/Q_{2,2} \\ x_5 + x_6 \\ x_7 + x_8 \\ g(x_1) - g(x_2) + r_{1,1}x_5 - r_{2,1}x_6 \\ g(x_3) - g(x_4) + r_{1,2}x_7 - r_{2,2}x_8 \end{bmatrix}, \quad u(t) = \begin{bmatrix} 0 \\ 0 \\ 0 \\ 0 \\ I \\ I \\ 0 \\ 0 \end{bmatrix},
\end{align}
and $H(x)$ once again takes the same form as that in \eqref{H}. Jacobians in (S3), (S4), and (O1) can be computed by automatic differentiation \cite{terrell1997observability,campbell1994utilization}. It is also confirmed numerically that the uniform differentiation index is $\gamma = 1$. The objective is to find the least indices $\gamma$ and $\delta$ that would render the battery pack system under minimal sensing smoothly observable. Numerically, utilizing the output function \eqref{H}, the least $\gamma$ and $\delta$ for the DAE system \eqref{DAE_dyn}-\eqref{DAE_out} to be smoothly observable are $\gamma = 3$ and $\delta = 3$. It is noted that up to $\ell$-th order gradients of $g(x)$ must be checked, where $\ell = \max\{(\gamma+1),(\delta+1)\}$. For condition (O1) to be satisfied, $g^{(i)}(x)$ with $i = \{1,2,\cdots,\ell\}$ should not be zero simultaneously. As can be seen from the 2P2S case, up to 4-th order gradients of OCV are involved. It has been studied extensively in the literature, e.g., \cite{Lin-2015,moura2016-electrolyte}, that higher order gradients of OCV approach zero, in particular in the middle SOC range (around 15\%-90\%). Hence, when only the pack total voltage is measured, the observability of individual cell SOC is expected to be weak. Nonetheless, at high and low SOC ranges, the OCV function $g$ is generally highly nonlinear with respect to SOC, rendering significant high-order gradients $g^{(i)}(x)$, $i = \{1,2,\cdots,\ell\}$. As has been previously highlighted in \cite{Lin-2015}, the high and low SOC ends are the regions where the risks of over-charge and over-discharge are critical, and high gradients in these regions should facilitate notable individual cell SOC observability from only pack voltage measurements. In addition, since the cells in the pack are heterogeneous, it is anticipated that cells would not have identical SOC values at any given time instant $t$, which ultimately enhances condition (O1) in Theorem \ref{SO}. However, it is also worth noting that in the extreme cases where all cells are identically parameterized, observability for individual cell SOCs will be completely lost as the cells are not distinguishable from one another. This trivial case just boils down to a single cell estimation problem.

The classical approach to study observability of a nonlinear battery syste, is to linearize the model, as done in Section \ref{sec:Linear_Obs} and e.g., \cite{ZhangDescriptor,Rausch-2013}. By doing so, the cell observability condition is only determined by the first-order derivative of OCV function. However, we can conclude from the analysis in Section \ref{sec:Linear_Obs} that this is conservative and less informative. Same conclusion was also drawn in \cite{Zhao-2017} through the local observability analysis of a single cell. 
The more detailed analysis based on smooth observability of the nonlinear battery pack DAE system showed that OCV gradients must be different than zero to guarantee local observability, but it does not need to be the first-order gradient. It relies on the fact that the observability condition (O1) explicitly depends on high-order gradients of OCV function, and it can be analytically obtained through e.g., symbolic software.

It is finally emphasized that in order for cell SOCs to be locally observable from only pack-level voltage and current measurements (i.e., minimal sensing), high-order gradients of OCV function must be checked. Although fewer measurement signals weaken cell-level observability, which is expected, our analysis provides considerable incentive to significantly reduce the number of sensors in a battery pack while maintaining enough cell-level observability.

\section{Design of State Observers}
\label{s:obs_design}

For the purpose of observer design, we investigated different notions of observability in Section \ref{s:obs_ana} and discussed the conditions under which the battery pack system with minimal sensing is smoothly observable. These conditions essentially ensure that there exists a state observer for estimating the SOC and current of individual cells modeled by the battery pack system (\ref{pack_dyn})-(\ref{pack_out}). In this section, we propose a Luenberger type state observer to asymptotically estimate cell-level SOCs and currents when only pack-level voltage and current are available for measurement.

The following observer with linear output error injection is proposed for the battery pack plant model (\ref{pack_dyn})-(\ref{pack_out}):
\begin{align}
    \label{obs_dyn}
    E\dot{\hat{x}}(t) & = A\hat{x}(t) + \Phi(\hat{x}(t),I(t)) + K\left[y(t)-C\hat{x}(t)-h(\hat{x}(t))\right], \\
    \label{obs_out}
    \hat{y}(t) & = C\hat{x}(t) + h(\hat{x}(t)),
\end{align}
where $\hat{x}(t)$ indicates the estimation of $x(t)$, and $K \in \mathbb{R}^{n}$ is the observer gain vector to be designed such that the differential states (individual cell SOCs) and the algebraic states (cell local currents) converge to the truth states asymptotically, i.e., $\hat{x}(t)\rightarrow x(t)$ as $t \rightarrow \infty$. Moreover, $K = [K_s^\top \quad K_u^\top]^\top$ with $K_s \in \mathbb{R}^N$ and $K_u \in \mathbb{R}^N$ (recall that $n = 2N$ and $N = N_p \times N_s$). The above observer structure adopts the linear output error injection method \cite{rajamani1998observers,kaprielian1992observer}, although the battery pack plant model is nonlinear. Theorem \ref{thm_obs} establishes the convergence properties of the proposed observer.

\begin{thm} \label{thm_obs}
Consider the battery pack plant model dynamics \eqref{pack_dyn}-\eqref{pack_out}, and suppose the matrix $[A_{22} \quad C]^\top$ has rank $N$. Let
\begin{equation}
    G = (A-KH) = \begin{bmatrix} G_{11} & G_{12} \\ G_{21} & G_{22} \end{bmatrix},
\end{equation}
and define the matrix
\begin{equation}
    \widetilde{G} = (G_{11}-G_{12}G_{22}^{-1}G_{21}).
\end{equation}
Suppose the function
\begin{equation}
    L(x,I) = \Phi_s(x)-G_{12}G_{22}^{-1}\Phi_u(x,I) + \left(G_{12}G_{22}^{-1}K_u-K_s\right)h(x),
\end{equation}
is Lipschitz continuous with respect to $x$, in which $\Phi = [\Phi_s \quad \Phi_u]^\top$ was defined in the vicinity of \eqref{xsu}. That is,
\begin{align}
    \|L(x_1,I) - L(x_2,I)\| & \leq \gamma \|x_1 - x_2\|
\end{align}
for any feasible $x_1,x_2 \in X$, where $\gamma \in \mathbb{R}$ is the Lipschitz constant. If the observer gain $K$ is chosen to ensure that $\widetilde{G}$ is stable, and
\begin{equation}
    \label{cond_sing}
    \min_{\omega \in \mathbb{R}^+} \sigma_{\rm min}\left(\tilde{G}-j\omega \mathbf{I}_{N \times N}\right) > \gamma,
\end{equation}
where $\sigma_{\rm min}(\cdot)$ denotes the minimum singular value. Then the zero equilibrium of the dynamics of estimation error  $e(t) = x(t) - \hat{x}(t)$ given by
\begin{equation}
    \label{error_dyn}
    E\dot{e} = Ge + \Phi(x,I) - \Phi(\hat{x},I) - K\left[h(x) - h(\hat{x})\right]
\end{equation}
is asymptotically convergent to zero.
\end{thm}

\begin{pf}
Let the state estimation error $e = [e_s^\top \quad e_u^\top]^\top$, with $e_s = s - \hat{s}$ being the estimation error for the differential states and $e_u = u - \hat{u}$ the estimation error for the algebraic states. Then (\ref{error_dyn}) can be decomposed into
\begin{align}
    \label{error_sys}
    \begin{bmatrix} \mathbf{I}_{N \times N} & \mathbf{0} \\ \mathbf{0} & \mathbf{0} \end{bmatrix} \begin{bmatrix} \dot{e}_s \\ \dot{e}_u  \end{bmatrix} = & \begin{bmatrix} G_{11} & G_{12} \\ G_{21} & G_{22} \end{bmatrix} \begin{bmatrix} e_s \\ e_u \end{bmatrix} + \begin{bmatrix} \Phi_s(x) - \Phi_s(\hat{x}) \\ \Phi_u(x,I) - \Phi_u(\hat{x},I)  \end{bmatrix} \nonumber \\
    & - \begin{bmatrix} K_s \\ K_u \end{bmatrix}\left[Cx - C\hat{x} + h(x) - h(\hat{x})\right].
\end{align}
We highlight that $G_{22}$ can be non-singular (i.e., invertible) if the linear part of (\ref{pack_dyn}) is impulse observable \cite{kaprielian1992observer}, i.e., the matrix $[A_{22} \quad C]^\top$ has rank $N$. Then the estimation error system (\ref{error_sys}) is equivalently described by
\begin{align}
    \label{ex_dyn}
    \dot{e}_s = {} & \left(G_{11}-G_{12}G_{22}^{-1}G_{21}\right)e_s \nonumber \\
    & + \left[\Phi_s(x)-G_{12}G_{22}^{-1}\Phi_u(x,I)\right] + \left(G_{12}G_{22}^{-1}K_u-K_s\right)h(x) \nonumber \\
    & - \left[\Phi_s(\hat{x})-G_{12}G_{22}^{-1}\Phi_u(\hat{x},I)\right]  - \left(G_{12}G_{22}^{-1}K_u-K_s\right)h(\hat{x}) \nonumber \\
    = {} & \widetilde{G}e_s + L(x) - L(\hat{x}),
\end{align}
along with the algebraic equation
\begin{align}
    \label{eu_alg}
    e_u = -G_{22}^{-1}G_{21}e_s - G_{22}^{-1}\left[\Phi_u(x) - \Phi_u(\hat{x})\right] + G_{22}^{-1}K_u\left[h(x) - h(\hat{x})\right].
\end{align}

Consider the following Lyapunov function for the error dynamics (\ref{ex_dyn}), corresponding to the differential states $e_s$,
\begin{equation}
    \label{Lyap}
    W(t) = \frac{1}{2}e_s^\top P e_s, \quad P = P^\top \succ 0.
\end{equation}
The time derivative of the Lyapunov function $W(t)$ along the trajectory of $e_s$ is computed by
\begin{align}
    \label{W_dot}
    \dot{W}(t) = {} & \frac{1}{2}\dot{e}_s^\top P e_s + \frac{1}{2}e_s^\top P \dot{e}_s \nonumber \\
    = {} & \frac{1}{2}e_s^\top(\widetilde{G}^\top P+P\widetilde{G})e_s + e_s^\top P [L(x) - L(\hat{x})] \nonumber \\
    \leq {} & \frac{1}{2}e_s^\top (\widetilde{G}^\top P+P\widetilde{G}) e_s + \|Pe_s\|\cdot\|L(x) - L(\hat{x})\| \nonumber \\
    \leq {} & \frac{1}{2}e_s^\top (\widetilde{G}^\top P+P\widetilde{G}) e_s + \gamma\|Pe_s\|\cdot\|e_s\| \nonumber \\
    \leq {} & \frac{1}{2}e_s^\top\left[\widetilde{G}^\top P+P\widetilde{G}+\gamma^2PP+\mathbf{I}_{N \times N}\right]e_s,
\end{align}
where the inequality
\begin{equation}
    2\gamma\|Pe_s\|\cdot\|e_s\| \leq \gamma^2e_s^\top PP e_s + e_s^\top e_s
\end{equation}
has been utilized in the last inequality of \eqref{W_dot}. According to Theorem 2 in \cite{rajamani1998observers}, if $\tilde{G}$ is stable and $\min_{\omega\in\mathbb{R}^+}\sigma_{\rm min}(\tilde{G}-j\omega \mathbf{I}_{N \times N}) > \gamma$, then there exists $\varepsilon > 0$ and $P = P^\top \succ 0$ such that
\begin{equation}
    \label{Ric_eq}
    \tilde{G}^\top P + P \tilde{G} + \gamma^2 P P + \mathbf{I}_{N \times N} + \varepsilon \mathbf{I}_{N \times N} = 0.
\end{equation}
Therefore, in view of \eqref{W_dot} and \eqref{Ric_eq}, $\dot{W}(t) < 0$, and the estimation error $e_s$ is asymptotically stable. Under this scenario, when $t \rightarrow \infty$, [$\Phi_u(x) - \Phi_u(\hat{x})] \rightarrow \mathbf{0}$, and $[h(x) - h(\hat{x})] \rightarrow 0$. Hence, according to \eqref{eu_alg}, the estimation error $e_u$ for the algebraic states also converges to zero asymptotically. This completes the proof.
\end{pf}

\begin{rmk}
Note that the first $N$ columns in matrix $A$ and matrix $C$ are zero columns. See, for example, matrix $A$ in \eqref{H} for the 2P2S configuration. In consequence, $G_{11}$ and $G_{21}$ are zero matrices, which yields a marginally stable $\tilde{G}$ regardless of the choice of observer gain matrix $K$. This violates the observer design conditions in Theorem \ref{thm_obs}. However, we observe that the matrix $A$ can essentially be freely assigned by adding stabilizing terms in the first $N$ columns, and the added terms are cancelled through the nonlinear term $\Phi$, which could potentially alter the Lipschitz constant $\gamma$.
\end{rmk}

\begin{rmk}
The Lipschitz constant $\gamma$ could be obtained by computing the infinity norm of function $L(x)$ with respect to the state $x$, i.e., $\gamma = \|\partial L/\partial x\|_\infty$. In addition, condition \eqref{cond_sing} can be practically verified if 
\begin{equation}
    {\rm Re}(-\lambda) > \kappa(R)\gamma,
\end{equation}
where $\lambda$ is the eigenvalues of $\tilde{G}$ and $\kappa(R)$ is the condition number of matrix $R$, in which $\tilde{G} = R \Lambda R^{-1}$. Interested readers may refer to Theorem 5 in \cite{rajamani1998observers} for more details and proof.
\end{rmk}

In view of Remark \ref{DAE_RED}, the observer \eqref{obs_dyn}-\eqref{obs_out} is proposed directly on the nonlinear battery pack DAE model without any model reductions. It has been established in Theorem \ref{thm_obs} that both differential and algebraic states are estimated in a feedback fashion with asymptotic convergence. Ultimately, the individual cell SOCs and currents can be effectively estimated with guaranteed mathematical convergence using only pack-level voltage and current measurements.

\section{Simulations} \label{s:sim}

\begin{figure}[t]
	\centering
	\includegraphics[trim = 0mm 0mm 0mm 6mm, clip, width=0.7\linewidth]{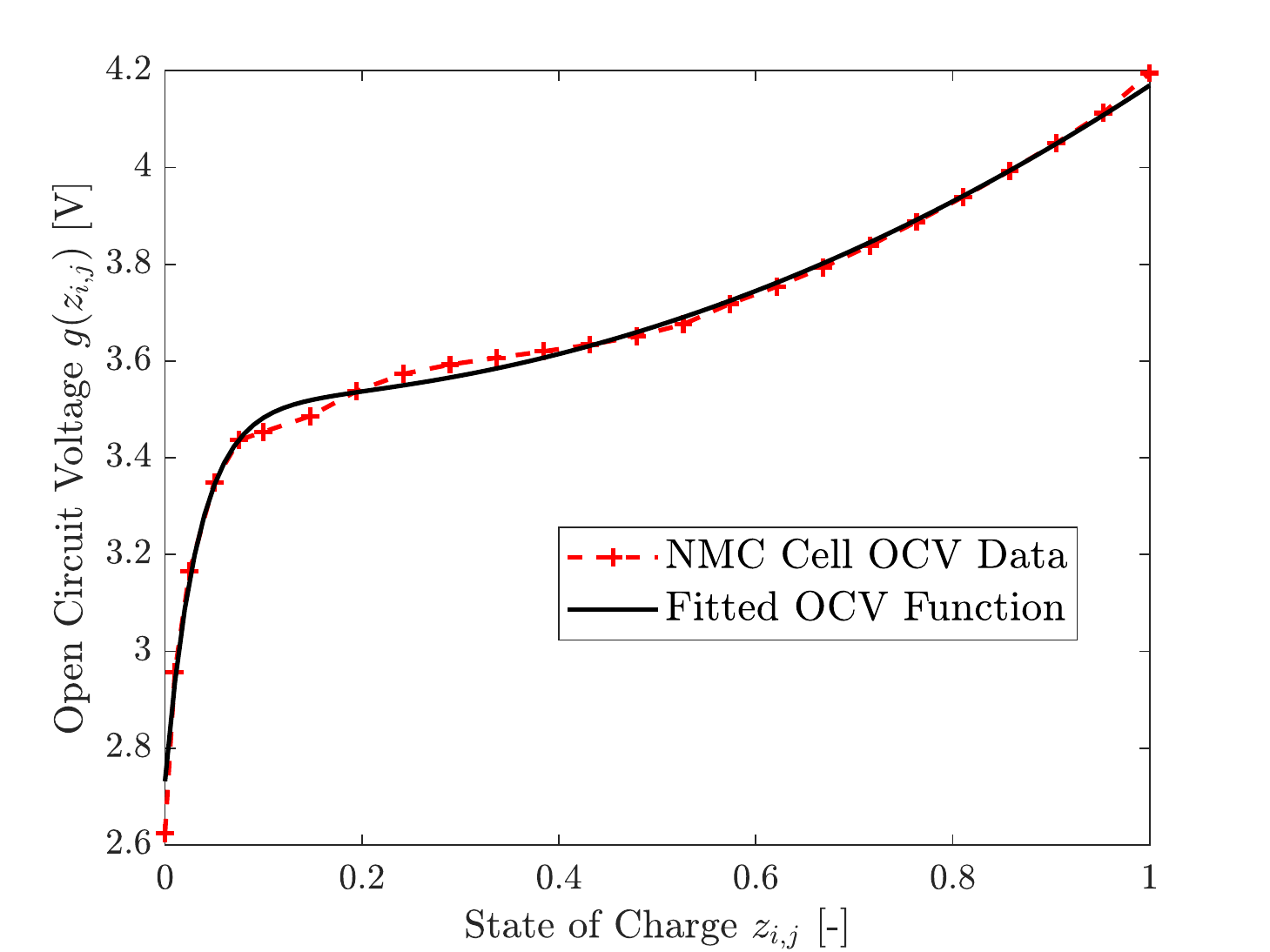}
	\caption{Open circuit voltage for a Graphite/NMC cells. The fitted OCV functional form is $g(z_{i,j}) = p_1 e^{(\alpha_1 z_{i,j})} + p_2 e^{(\alpha_2 z_{i,j})} + p_3z_{i,j}^2$ adopted from \cite{improved_OCV}.}
	\label{fig:OCV}
\end{figure}

This section presents a simulation study that demonstrates the performance of the proposed state observer for individual cell SOC estimation in a battery pack under minimal sensing scenario. For the ease of presentation and without loss of generality, the numerical implementation is conducted for the 2P2S case. All cells in the pack are of Graphite/NMC type, whose open circuit voltage is shown in Figure \ref{fig:OCV}. We consider the situation in which the cells may differ in their initial SOCs and model parameters, but subject to the same SOC-OCV relationship. The assumption that SOC-OCV relationship is the same is based on the fact that this is a thermodynamic property and only a function of the electrode materials.  Thus, processing variation at the material level is likely to not impact the SOC-OCV variation, compared to other quantities such as capacity, resistance, etc. The numerical values for the model parameters are enumerated in Table \ref{params}. The considered setup guarantees local smooth observability based on the analysis in Section \ref{smooth_obs}. For all simulations, the state estimates are intentionally initialized at incorrect values. The true SOC initial conditions are $20\%$, $25\%$, $15\%$, and $22\%$. The observer initial conditions are perturbed by 50\% of the true values, and they are given by $30\%$, $37.5\%$, $22.5\%$, and $33\%$ (Table \ref{params}). In the presented simulations, we utilize the battery pack plant model simulated data to validate the proposed observer. Ultimately, the DAE plant model and the corresponding observer system are solved using the publicly available numerical solvers in MATLAB\textsuperscript{\textregistered}. A crucial step in the numerical integration is to compute consistent initial conditions.

\begin{table}[t]
\caption{Model Parameters in Simulations} \label{params}
\centering
\begin{tabular}{@{}cccccl@{}}
\toprule
 & \multicolumn{2}{c}{Module 1} & \multicolumn{2}{c}{Module 2} &       \\ \midrule
 & Cell 1 & Cell2 & Cell 1 & Cell 2 & Units \\ \midrule
$r$ & 0.1 & 0.22 & 0.3 & 0.13 & [$\Omega$] \\ \midrule
$Q$ & 1500 & 1800 & 1200 & 2000 & [A$\cdot$sec] \\ \midrule
$z_0$ & 0.2 & 0.25 & 0.15 & 0.22 & [--] \\ \midrule
$\hat{z}_0$ & 0.3 & 0.375 & 0.225 & 0.33 & [--] \\ \bottomrule
\end{tabular}
\end{table}

\begin{figure}[t]
	\centering
	\includegraphics[trim = 6.3mm 0mm 10mm 4mm, clip, width=0.7\linewidth]{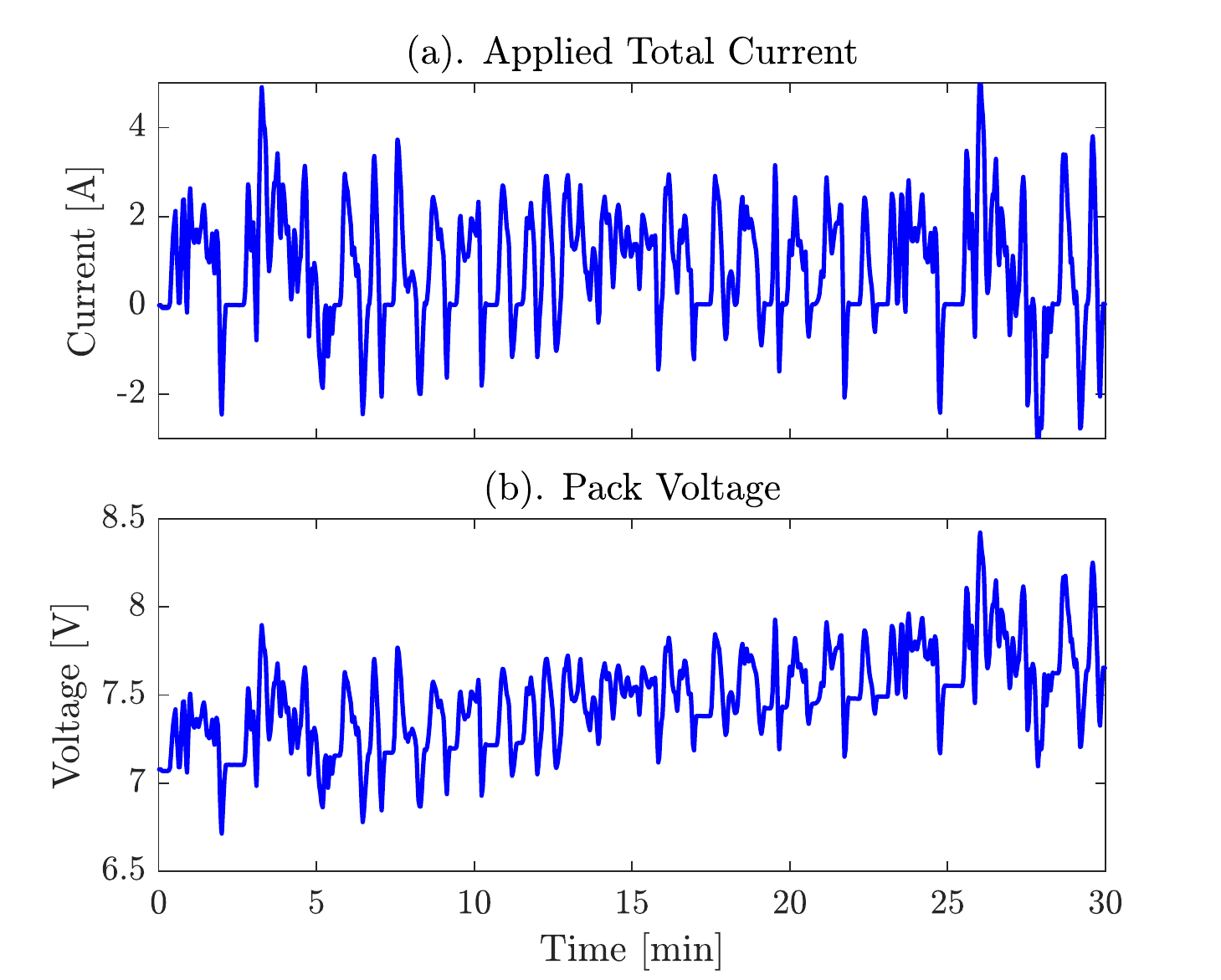}
	\caption{Plant model simulation for 2P2S with model parameters listed in Table \ref{params}. (a) applied total voltage; (b) pack level voltage measurement.}
	\label{fig:Plant}
\end{figure}

\begin{figure}[t]
	\centering
	\includegraphics[trim = 4mm 0mm 10mm 0mm, clip, width=0.7\linewidth]{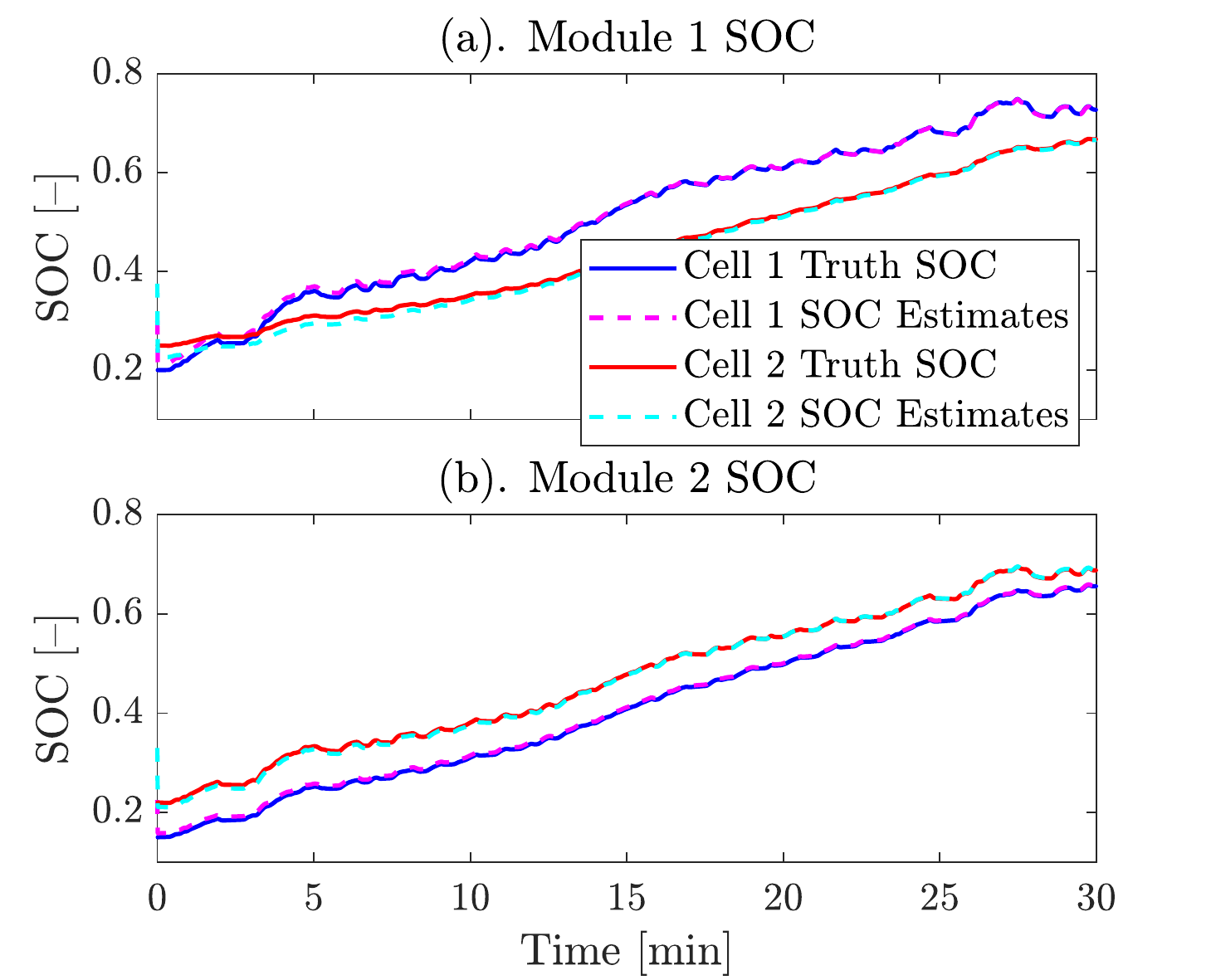}
	\caption{Individual cell SOC estimation performance. SOC trajectories will not synchronize when cells are heterogeneous. All SOC estimates are initialized with 50\% errors, then quickly converge to the truth SOCs asymptotically using pack-level voltage measurement only. (a) parallel module 1 cell SOC estimates; (b) parallel module 2 cell SOC estimates.}
	\label{fig:SOC}
\end{figure}

\begin{figure}[t]
	\centering
	\includegraphics[trim = 7mm 8mm 10mm 4mm, clip, width=0.69\linewidth]{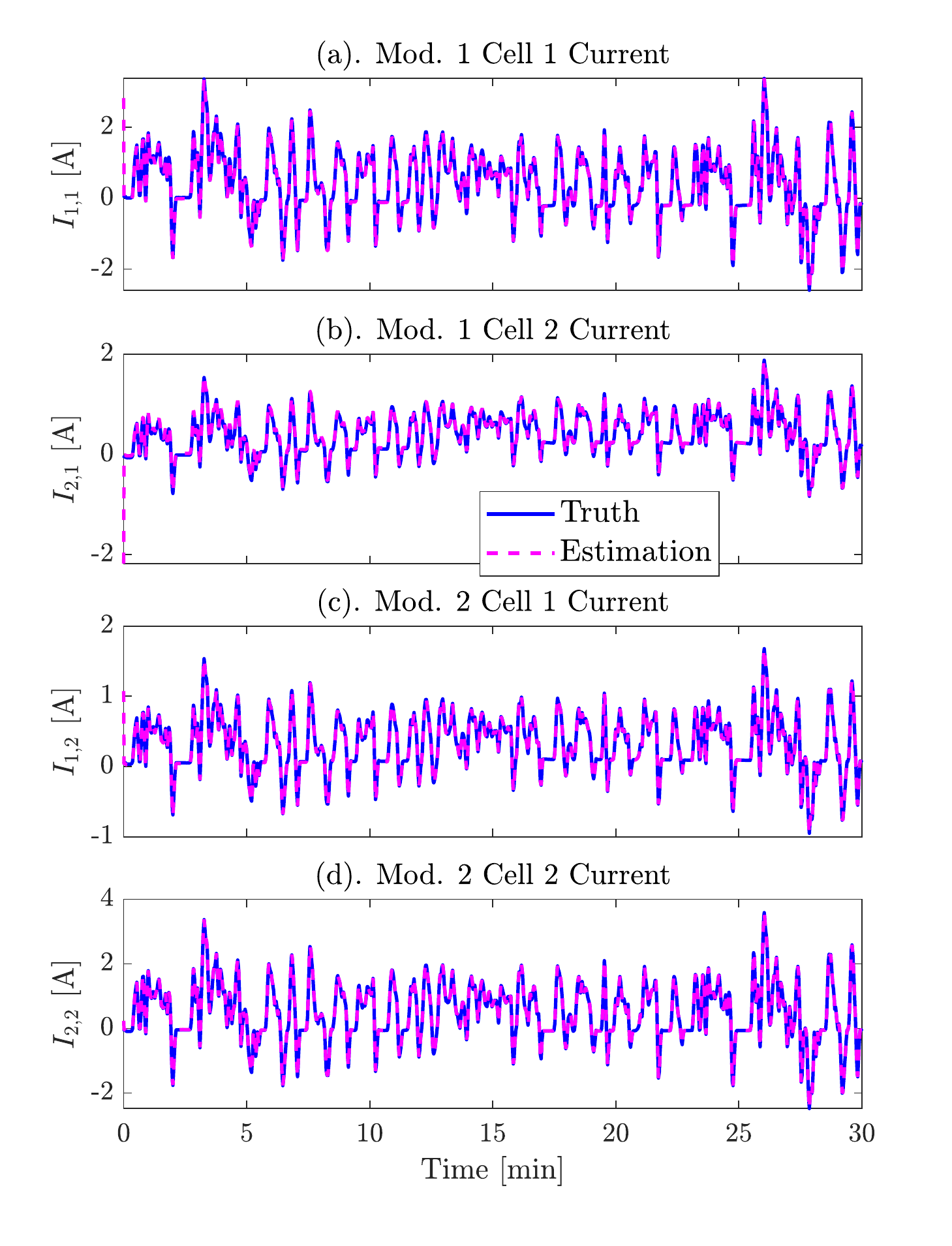}
	\caption{Individual cell current estimation performance. Cell current estimates are initialized far from the truth, for the purpose of consistent initial conditions. These algebraic state estimates converge instantly. (a) module 1 cell 1 current estimates; (b) module 1 cell 2 current estimates; (c) module 2 cell 1 current estimates; (d) module 2 cell 2 current estimates.}
	\label{fig:Current}
\end{figure}

In this simulation study, the total applied current at the pack level is appropriately scaled from an Urban Dynamometer Driving Schedule (UDDS) drive cycle to emulate a practical electric vehicle driving pattern, and it is sketched in Figure \ref{fig:Plant}(a). The simulated (and measured) pack voltage responses are reported in Figure \ref{fig:Plant}(b). The objective is to reconstruct individual cell SOC signals asymptotically from the pack-level current and voltage measurements only, via observer gain selection according to Theorem \ref{thm_obs}.
The observer in (\ref{obs_dyn})-(\ref{obs_out}) is used to estimate the individual cell SOCs and the cell local currents. 

Figure \ref{fig:SOC} and Figure \ref{fig:Current} demonstrate the estimation performance. These results are numerically generated using the observer gain
\begin{equation}
K =
    \begin{bmatrix}
    0.65 & 1.09 & 0.80 & 0.83 & 0.36 & 0.51 & 0.04 & 1.85 \\
    1.31 & 1.87 & 0.33 & 1.84 & 1.59 & 1.15 & 0.88 & 0.52 \\
    1.50 & 0.46 & 0.13 & 1.53 & 1.34 & 1.43 & 1.28 & 0.84
    \end{bmatrix}^\top, \nonumber
\end{equation} 
which satisfies the observer conditions in Theorem \ref{thm_obs}. Figure \ref{fig:SOC} illustrates the convergence behavior of individual cell SOCs. The solid curves are the truth SOCs that are simulated from the plant model, and the dashed curves represent the SOC estimates. Notice again that when cells are heterogeneous, their SOC trajectories will not synchronize, even within a parallel module that self balances voltages. Despite significantly incorrect initial conditions (50\% initial errors for all cases), the SOC estimates quickly converge to the truth values. After a rapid initial transient, the SOC estimates have root mean squared (RMS) errors of 0.13\%, 0.37\%, 0.25\%, and 0.092\% for module 1 cell 1, module 1 cell 2, module 2 cell 1, and module 2 cell 2, respectively. Furthermore, Figure \ref{fig:Current} portrays the estimates for the algebraic states (cell local currents). Note that the current estimates are initialized considerably far from the real initial spots of the truth currents (see the dashed magenta curves in Figure \ref{fig:Current}). This is because the initial conditions of algebraic state estimates are calculated based on differential states' initial conditions to form consistent initial conditions for the observer DAE system \eqref{obs_dyn}-\eqref{obs_out}. Drastic SOC initial estimation errors induce enormous errors in cell local current estimates. Despite large initial errors, the algebraic state estimates are able to recover the truth signals almost instantly. The performances of the observer confirms the asymptotic zero error convergence conclusions from Theorem \ref{thm_obs}.

\section{Conclusion} \label{s:conclusion}
In this study, the cell-level SOC estimation problem in a battery pack is investigated. In contrast to conventional approaches where a battery pack is represented by a lumped cell in which cell-level information is ignored, the framework proposed in this paper rigorously exploits the high fidelity cell-by-cell resolution using the interconnection of single cell models. We further challenge the problem set-up with minimal sensing scenario, where the pack-level voltage and current are the only measurable signals. It is shown that non-zero high-order gradients of OCV function is required for the smooth observability of cell-level SOCs and currents, whereas the observability analysis from linearizing the nonlinear battery pack model does not provide conclusive results.

A nonlinear DAE system has been proposed to model the battery pack with parallel-series arrangements, and a state observer for such a DAE system has been developed. 
This modeling framework fits naturally with battery applications, given the interconnections arising from Kirchhoff's laws. 
The design procedure used to build the state observer from this model avoids linearization or canonical transformations, and it only relies on the assumption of Lipschitz nonlinearities. 
The resulting state observer benefits from considering the unknown cell-level currents as algebraic states to be simultaneously estimated with the differential states in a feedback fashion. The effectiveness of the proposed estimation approach was demonstrated in simulation.

Consequently it is noted that although fewer measurement signals weaken cell-level observability, the analysis in this paper provides considerable incentives to significantly reduce the number of sensors in a battery pack. Future work will explore the effects of temperature \cite{zhang2021thermint} on cell-level state estimation with reduced sensing.

\section*{Acknowledgments}

Luis D. Couto would like to thank the Wiener-Anspach Foundation for its financial support.

\bibliography{main}

\end{document}